\documentclass[aps,prb,reprint,floatfix,superscriptaddress]{revtex4-1}
\usepackage[english]{babel}
\usepackage[T1]{fontenc}
\usepackage{bbold}
\usepackage{epsfig}
\usepackage{amsmath}
\usepackage{placeins}
\usepackage{hyperref}

\usepackage{subcaption}

\usepackage{array}
\newcolumntype{C}[1]{>{\centering\arraybackslash}m{#1}}
\usepackage{overpic}

\begin{document}

\title{Collective effects in tilted Weyl cones:\\ optical conductivity, polarization, and Coulomb interactions reshaping the cone}
\author{Fabrizio Detassis}
\affiliation{Institute for Theoretical Physics and Center for Extreme Matter and Emergent Phenomena, Utrecht University, Leuvenlaan 4, 3584 CE Utrecht, The Netherlands}
\author{Lars Fritz}
\affiliation{Institute for Theoretical Physics and Center for Extreme Matter and Emergent Phenomena, Utrecht University, Leuvenlaan 4, 3584 CE Utrecht, The Netherlands}
\author{Simonas Grubinskas}
\affiliation{Institute for Theoretical Physics and Center for Extreme Matter and Emergent Phenomena, Utrecht University, Leuvenlaan 4, 3584 CE Utrecht, The Netherlands}

\begin{abstract}
Recently, the existence of Dirac/Weyl cones in three dimensional systems has been demonstrated experimentally. While in high energy physics the isotropy of the Dirac/Weyl cones is guaranteed by relativistic invariance, in condensed matter systems corrections to this can occur, one possible type being a tilt. 
In this paper we study the effect of of tilted Weyl cones in collective effects. We study both the optical conductivity as well as the polarization function. We also investigate the perturbative effect of long-range Coulomb interactions using a renormalization group calculation. We find that the tilt is perturbatively renormalized towards zero and at low energies the system flows to an effectively untilted theory.  
\end{abstract}
\maketitle

\section{Introduction}

Dirac semimetals (DSMs) and Weyl semimetals (WSMs) are electronic states of matter characterized by the existence of three-dimensional Dirac nodes\cite{wan2011topological,balents2011viewpoint,weng2015weyl,huang2015weyl,turner2013beyond,Lv2015Experimental,xu2015discovery,xu2015discovery2,yang2015weyl,lv2015observation,xu2015observation,shekhar2015extremely}. Unlike in three dimensional (3D) DSMs, Weyl nodes of opposing chirality in WSMs are separated in momentum space. As such, they act as sources and sinks of Berry curvature.\cite{wan2011topological} WSMs have been likened\cite{yang2015weyl} to ``3D graphene'', but also exhibit a range of fascinating properties not observed in graphene or other topological semimetallic systems -- such as the chiral anomaly in transport\cite{zyuzin2012topological,liu2013chiral,zhang2015observation,PhysRevB.92.125141} and the appearance of open surface Fermi arcs in photoemission measurements.\cite{Lv2015Experimental,xu2015discovery,xu2015discovery2,yang2015weyl,lv2015observation,xu2015observation} The latter are the result of topologically-protected chiral states connecting bulk Weyl nodes of opposite chirality, projected onto the surface. Information about the bulk topology can therefore be obtained from the surface-projected Fermi arcs. 
Such topological properties are predicted to be robust to weak perturbations, including disorder from dilute impurities,\cite{bera2015dirty} since the Weyl nodes can only be annihilated in pairs of opposite chirality.\cite{wan2011topological}

A family of WSMs was predicted from band structure calculations in the monopnictide class\cite{weng2015weyl,huang2015weyl} -- and very recently Weyl fermion states have been discovered experimentally in the noncentrosymmetric (but time-reversal invariant) materials TaAs, NbAs, TaP and NbP.\cite{Lv2015Experimental,xu2015discovery,xu2015discovery2,yang2015weyl,lv2015observation,xu2015observation,shekhar2015extremely} In particular, surface Fermi arcs have been observed in angle-resolved photoemission spectroscopy (ARPES) experiments carried out for these systems.

More recently, it was appreciated that condensed matter Weyl cones, in contrast to fundamentally relativistic Dirac/Weyl fermions, do not have to be perfectly isotropic. They can be either slightly tilted~\cite{Bergholtz2015} or even tipped~\cite{soluyanov2015typeii}, in the latter case leading to electron and hole Fermi surfaces coexisting with the Weyl touching points. The  latter systems are commonly referred to as Weyl fermions type II~\cite{soluyanov2015typeii}. Obviously, this can lead to considerably more complicated situations for instance if disorder is present and in principle nodal quasiparticles can scatter resonantly to electron or hole Fermi pockets.  

In this paper we investigate the effect of a finite tilt in Weyl systems on the optical conductivity (here our result is also valid for type II) as well as the polarization. Our main finding is that the optical conductivity to a large extent is not altered by the existence of a finite tilt and does not show a strong directional dependence; the signature of the tilt vanishes completely in the extreme optical limit, {\it i.e.}, $\frac{\omega}{T} \gg 1$. The polarization, on the other hand, shows directional dependence. We furthermore study the effects of Coulomb interactions in such systems and contrast them from the standard isotropic case. Our main finding is that asymptotically the system recovers isotropy in the low-energy limit. 

The organization of the paper is as follows: we start with introducing the minimal model Sec.~\ref{sec:model}; from there we move towards the optical conductivity in Sec.~\ref{sec:optcon} and the polarization function in Sec.~\ref{sec:polfun}. Finally, we discuss the renormalization due to interactions in Sec.~\ref{sec:renorm} and finish with a discussion and conclusion in Sec.~\ref{sec:conc}.

\section{Minimal Model}\label{sec:model}
The starting point for our investigations is given by two tilted Weyl cone parametrized by the Bloch Hamiltonian
\begin{eqnarray}\label{eq:ham}
\mathcal{H}({{\bf{k}}})=v_F \tau_z \otimes \vec{\sigma} \cdot {{\bf{k}}}+v_F t  \tau_0 \otimes \sigma_0 {\bf{d}}\cdot {{\bf{k}}}
\end{eqnarray}
where ${\bf{d}}$ is a unit vector ($|{\bf{d}}|$=1), $t$ is a dimensionless parameter measuring the tilt, and $\sigma_0$ and $\tau_0$ are unit matrices while $\sigma^{x,y,z}$ and $\tau^{x,y,z}$ are the standard Pauli matrices. The eigenvalues are doubly degenerate and given by
\begin{eqnarray}
E_{\pm}({{{\bf{d}}}})=v_F t {{\bf{d}}} \cdot {\bf{k}} \pm v_F k \; .
\end{eqnarray}
The density of states remains semi-metallic as long as $t<1$ and becomes metallic for $t \ge 1$. A finite tilt term, {\it{i.e.}}, $t\neq 0$ cannot be added without breaking time reversal symmetry in a realistic model of two Dirac cones as can easily be shown by a symmetry analysis. The present implementation also breaks inversion symmetry; a version which does not break inversion symmetry is given by a tilt term of the form $v_F t  \tau_z \otimes \sigma_0 {{\bf{d}}}\cdot {\bf{k}}$ (note that this still breaks time reversal symmetry). In the following we assume that both Weyl points sit at the same energy. On the level of Eq.~\eqref{eq:ham} this requires time reversal symmetry breaking and intact inversion symmetry. However, also in the case of broken inversion symmetry and intact time reversal symmetry Weyl points can sit at the same energy, it only requires and even number of pairs and thus cannot be captured by Eq.~\eqref{eq:ham}.

\section{Optical Conductivity}\label{sec:optcon}

The optical conductivity is the response to an applied electric field which varies periodically in time. It is material characteristic and has been studied for untilted 3D Dirac/Weyl fermions extensively. In the presence of a finite tilt, the naive expectation is that there is a natural bias towards the tilt direction. The real part of the optical conductivity in general is given by
\begin{eqnarray}
\Re \; \sigma^{\alpha \beta} (\omega)=\frac{1}{\omega}\Im \; K^{\alpha \beta}(\omega^+,{\bf{q}}=0)\;.
\end{eqnarray}
In imaginary time formalism the linear response kernel for Hamiltonian, Eq.~\eqref{eq:ham}, in the absence of interactions and disorder, reads (see Appendix~\ref{app:optcon} for a derivation of the current vertex)
\begin{widetext}
\begin{eqnarray}\label{eq:kernel}
K^{\alpha \beta}(i\nu_n,{\bf{q}})=-\frac{e^2 v_F^2}{\beta}  \sum_{i \omega_n} \int \frac{d^3k}{(2\pi)^3}{\rm{tr}} \left[ \left(\tau_z \otimes \sigma^\alpha + t \tau_0 \otimes \sigma_0 d^\alpha \right)G(i\omega_n+i\nu_n,{{\bf{k}}}+{\bf{q}}) \left(\tau_z \otimes \sigma^\beta + t \tau_0 \otimes \sigma_0 d^\beta \right) G(i\omega_n,{{\bf{k}}})\right]  \nonumber \\
\end{eqnarray}
\end{widetext}
where $\omega_n=(2n+1)\pi T$, $\nu_n=2n\pi T$ ($n$ is integer), $\alpha=x,y,z$. We obtain $K^{\alpha \beta}(\omega^+,{\bf{q}})$ by means of analytic continuation via $i\nu_n \to \omega^+=\omega+i 0^+$. The Green function itself standardly is given by
\begin{eqnarray}
G(i\omega_n,{{\bf{k}}})=\left( i \omega_n - \mathcal{H}({\bf{k}}) \right)^{-1}\;.
\end{eqnarray}

We distinguish two response functions, $\sigma^\parallel$ and $\sigma^\perp$, where the former is the response if the electric field is applied along the tilting direction, while the latter corresponds to a field in the plane perpendicular to ${\bf{d}}$. 
After a series of manipulations (for details of the calculation see Appendix~\ref{app:optcon}) we can bring them into the compact form
\begin{eqnarray}
\sigma^\perp (\omega,T,t)&&=\frac{e^2 |\omega|}{32 v_F \pi} \int_{-1}^1 dx (1+x^2) \times \nonumber \\ &&\left(n_F\left(-\frac{|\omega|}{2}+\frac{|\omega| t x }{2}\right)-n_F\left(\frac{|\omega|}{2}+\frac{|\omega| t x }{2}\right) \right)\nonumber \\
\end{eqnarray}
and
\begin{eqnarray}
\sigma^\parallel (\omega,T,t)&&=\frac{e^2 |\omega|}{16 v_F \pi} \int_{-1}^1 dx (1-x^2) \times \nonumber \\ &&\left(n_F\left(-\frac{|\omega|}{2}+\frac{|\omega| t x }{2}\right)-n_F\left(\frac{|\omega|}{2}+\frac{|\omega| t x }{2}\right) \right) \;.\nonumber \\
\end{eqnarray}
Both integrals only depend on two dimensionless parameters, $\omega/T$ and $t$ (note $\hbar=k_B=1$).

If we assume that the electric field is applied in a direction which is at an angle $\phi$ with respect to the tilting vector $\bf{d}$, the resulting optical conductivity can be expressed as  
\begin{eqnarray}
\sigma(\omega,T,t,\phi)&=&\frac{1}{2}\left ( \sigma^\perp (\omega,T,t)+\sigma^\parallel (\omega,T,t)\right) \nonumber \\ &+& \frac{1}{2}\left(\sigma^\parallel (\omega,T,t)-\sigma^\perp (\omega,T,t)\right)\cos 2\phi \;.
\end{eqnarray}
 
%\begin{figure}
%\includegraphics[width=0.45\textwidth]{perp.pdf}
%\caption{This plot shows the difference $\sigma^\parallel (\omega,T,t)-\sigma^\perp (\omega,T,t)$ for values of $t=0.1,0.4,0.9,0.99$. }\label{fig:perp}
%\end{figure}

One interesting observation is that in the limit $\omega/T \gg 1$ the two components of the optical conductivity, $\sigma^\parallel (\omega,T,t)$ and $\sigma^\perp (\omega,T,t)$, do not depend on the tilt anymore.
To see this we note that in this limit we can approximate the Fermi function $n_F$ as a step function; consequently, the optical conductivities simplify and are given by
\begin{eqnarray}\label{eq:sigmaperp}
\sigma_\perp=\frac{e^2 |\omega|}{12 v_F \pi}\left( 1 -\theta (t-1) \left(1-\frac{3 t^2+1}{4 t^3} \right)\right)
\end{eqnarray}
and 
\begin{eqnarray}\label{eq:sigmaparallel}
\sigma_\parallel=\frac{e^2 |\omega|}{12 v_F \pi}\left( 1 -\theta (t-1) \left(1-\frac{3 t^2-1}{2 t^3} \right)\right) \;.
\end{eqnarray}
Note that this formula reproduces the result of the untilted case~\cite{hosur2012charge} for all values of $t<1$.
While this independence of the tilt sounds surprising, there is an intuitive picture. The optical conductivity probes the transitions of electrons from the valence to the conduction band caused by external electromagnetic radiation; it exchanges an energy $\omega$ and a momentum $\mathbf{k}=0$ with an electron in the valence band thereby exciting it into the conduction band.
 \begin{figure}[h]
    \includegraphics[width=0.4\columnwidth]{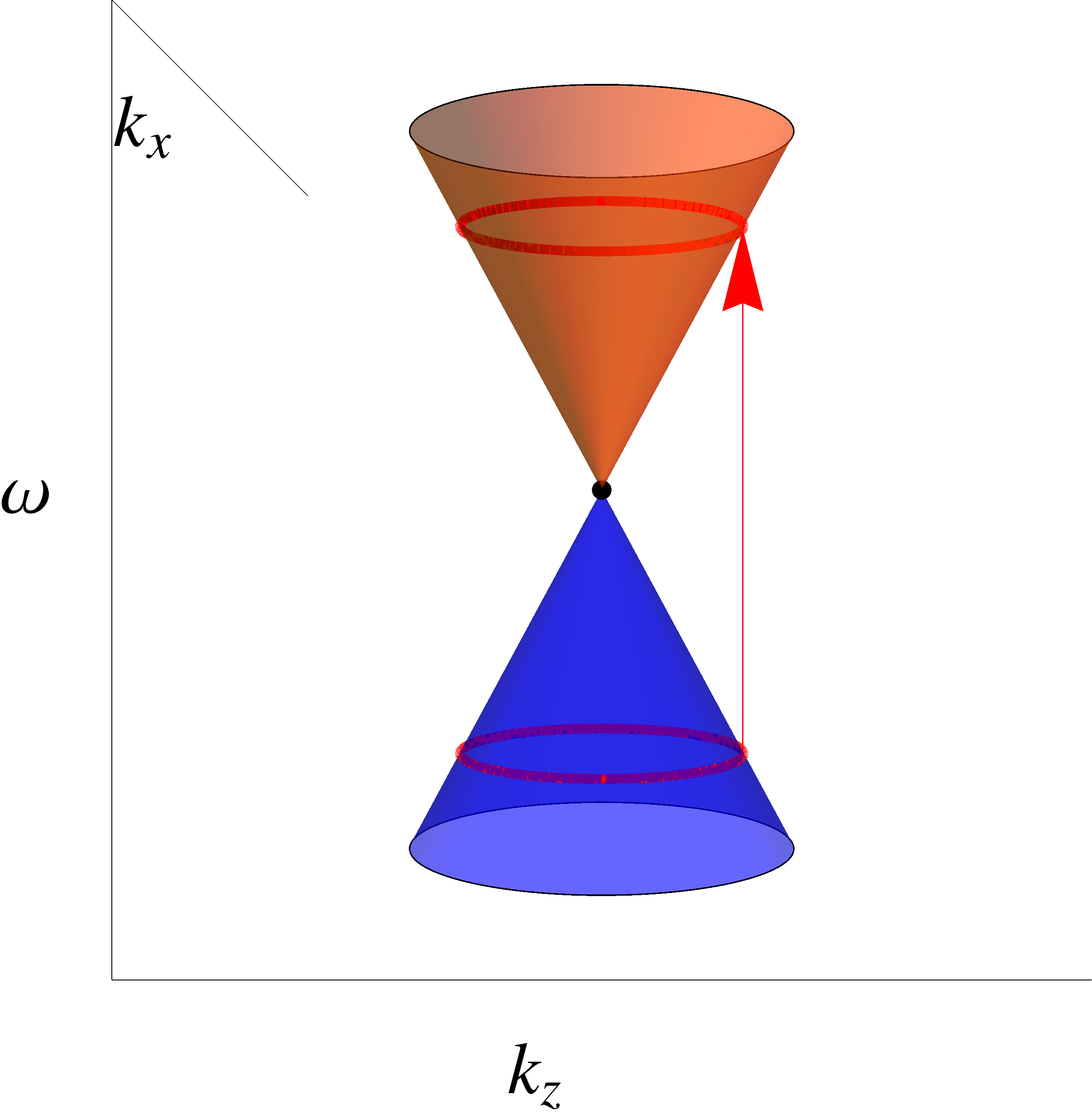}
    \hspace{1cm}
    \includegraphics[width=0.4\columnwidth]{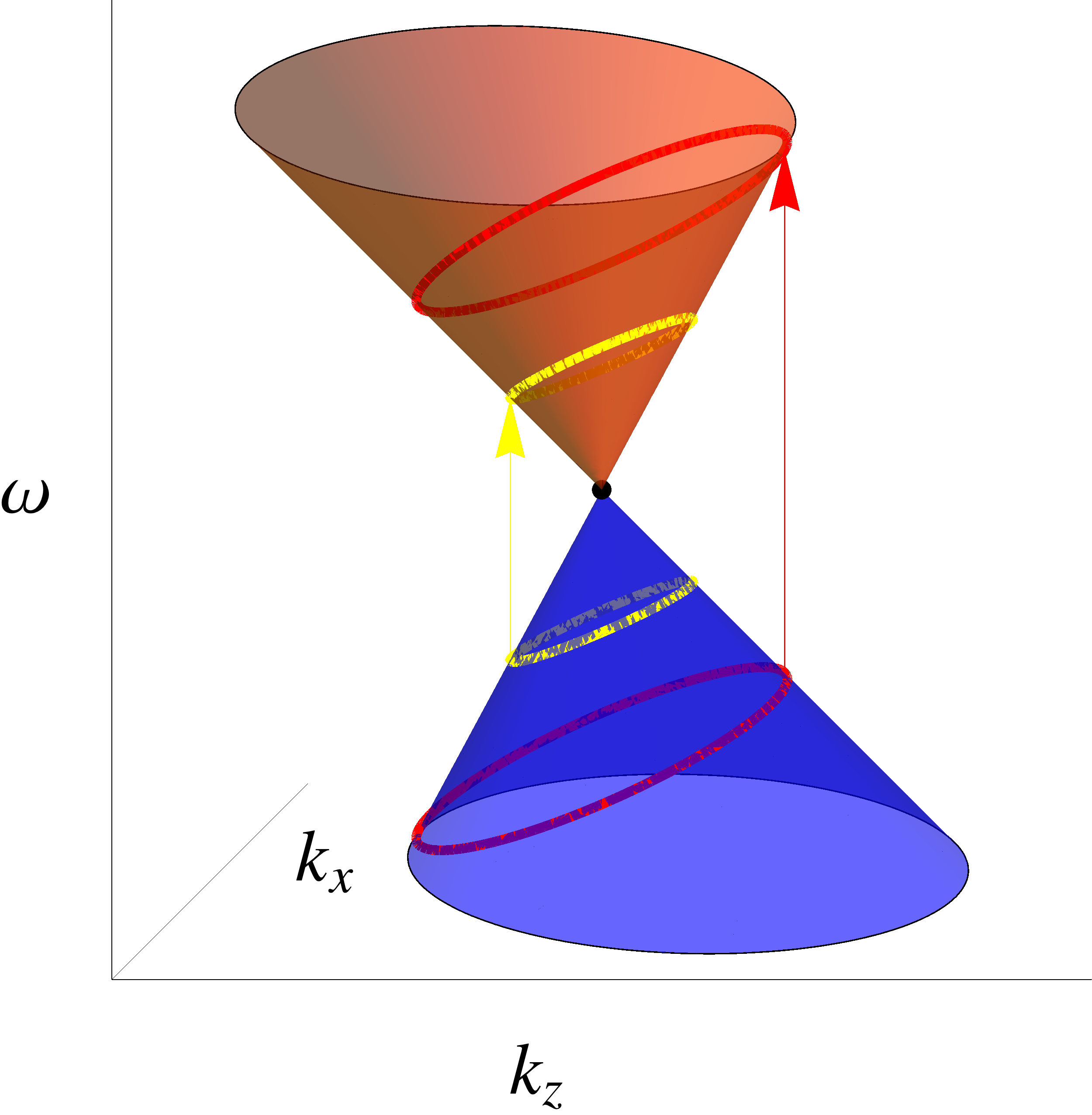}
    \hspace{1cm}
    \includegraphics[width=0.4\columnwidth]{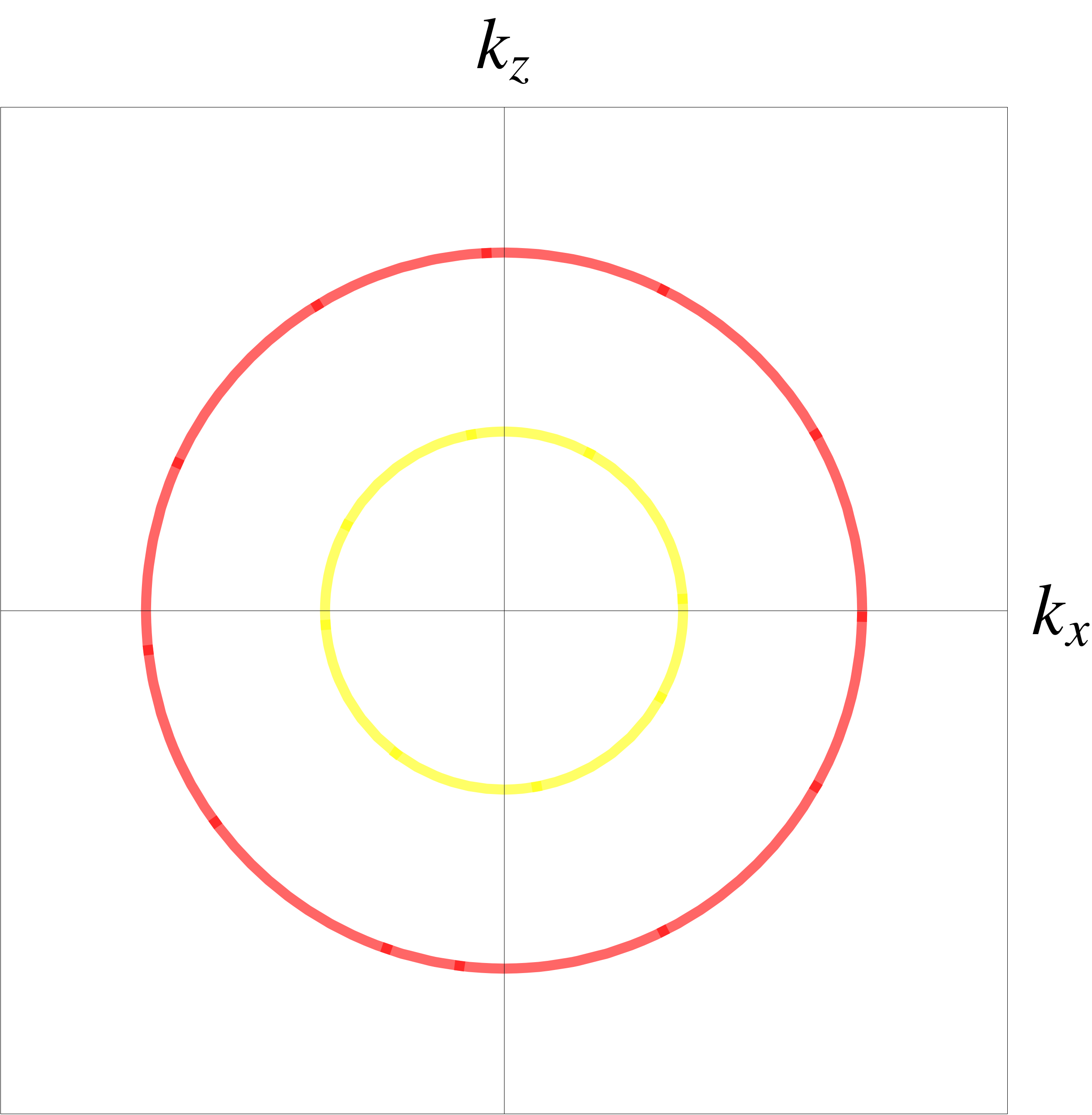}
    \caption{Graphical interpretation of the conductivity. Transitions at $\omega=2$ (red) and $\omega=1$ (yellow) and their projection on the $\omega=0$ plane.}\label{fig:gic}
  \end{figure} 
For untilted cones, the number of states that can undergo such a transitions are those lying on the circular region corresponding to an energy $E=-\omega/2$, which will get excited to the region at energy $E=\omega/2$. Since this process does not involve momentum transfer the transition corresponds to a vertical "jump". This essentially measures the density of states $\propto \omega^2$ integrated along a ring, leading to Eq.~\eqref{eq:sigmaperp} and Eq.~\eqref{eq:sigmaparallel}. \\
If we now consider the additional tilting term, the states that can be excited from valence to conduction band lie on an ellipsoid, tilted itself. The number of states is given by the line integral over the region of the density of states, and this latter changes along the ellipse. The two effects, length of the region and varying density of states, compensate each other in such a way that only the projection of the shaded areas onto the $\omega=0$ plane matters.\\
We conclude that the optical conductivity does not change as long as the Weyl cone is not tilted over, {\it i.e.}, as long as $t<1$ (even for general values of $\omega/T$ the conductivity is only very weakly
dependent on the tilt $t$). This also continues to be true if there is a finite chemical potential $\mu$. In that situation the optical conductivity is activated meaning there only is a response if the optical excitation frequency $\omega>\textrm{min}\left[\frac{2\mu}{1+t},\frac{2\mu}{1-t}\right]$ and it obeys Eq.~\eqref{eq:sigmaperp} and Eq.~\eqref{eq:sigmaparallel} for $\omega>\textrm{max}\left[\frac{2\mu}{1+t},\frac{2\mu}{1-t}\right]$. Inbetween it performs a smooth crossover.

\section{Polarization function}\label{sec:polfun}
The retarded polarization function $\pi_R (\omega,{\bf{q}})$ is given as the density-density correlation function~\cite{Altland}. At zero temperature the full expression for the polarization function reads
\begin{widetext}
\begin{eqnarray}\label{eq:pol}
\pi_R (\omega,{\bf{q}})&=&\frac{q^2}{24 \pi^2v_F} \ln \left(1+\frac{4\Lambda^2}{v_F^2 q^2-(\omega-t v_F {\bf{d}} \cdot {\bf{q}})^2} \right) \nonumber \\  &+& \frac{q^2}{24 \pi^2 v_F}\frac{8\Lambda^3}{\left( v_F^2 q^2-(\omega-t v_F {\bf{d}} \cdot {\bf{q}})^2\right)^{3/2}} \arctan \left(\frac{\sqrt{v_F^2 q^2-(\omega-t v_F {\bf{d}} \cdot {\bf{q}})^2}}{2\Lambda}-\frac{4\Lambda^2}{v_F^2 q^2-(\omega-t v_F {\bf{d}} \cdot {\bf{q}})^2} \right) 
\end{eqnarray}
\end{widetext} where $\Lambda$ corresponds to a momentum cutoff which has to be introduced to regularize the integral (details of the calculation are relegated to Appendix~\ref{app:polfun}).
The imaginary part which is related to the spectrum of the polarization is given by
\begin{eqnarray}
{\rm{Im}} \; \pi_R(\omega,{\bf{q}})=\frac{q^2}{24 \pi^2 v_F} \theta \left(\left(\omega-tv_F {\bf{d}}\cdot {\bf{q}}\right)^2-v_F^2 q^2 \right) \;.\nonumber \\
\end{eqnarray}
In contrast to the optical conductivity, this quantity shows directional dependence even at zero temperature. This can be traced back to the fact that we are looking for a response at finite external momentum. 
We can again decompose the spectrum into a perpendicular and parallel component given by
\begin{eqnarray}
{\rm{Im}} \; \pi^\perp_R(\omega,{\bf{q}})&=& \frac{q^2}{24 \pi^2 v_F} \theta \left( \omega^2-v_F^2 q^2 \right) \;, \nonumber \\
{\rm{Im}} \; \pi^\parallel_R(\omega,q)&=&\frac{q^2}{24 \pi^2 v_F} \theta \left(\left(\omega-t v_F q\right)^2-v_F^2 q^2 \right)\;.
\end{eqnarray}
Just like in the case of the optical conductivity we reproduce the zero-tilt result known from the literature~\cite{}.

\section{Tilted Weyl cones and Coulomb interaction}\label{sec:renorm}

Coulomb interactions in graphene are marginally irrelevant from a perturbative point of view~\cite{gonzalez1994,Son2007}. Their main effect is to renormalize the Fermi velocity upwards upon decreasing the energy scale. This picture has recently seen confirmation in experiments~\cite{elias2011}. However, from a strong coupling perspective there is room for a phase with broken symmetry and a dynamically generated excitonic gap~\cite{Khveshchenko2001, Son2007, Drut2009} and currently experiments are currently trying to find the so-called chiral symmetry breaking in ever cleaner samples.  

Three dimensional Dirac theories are fundamentally different in the sense that in the massless case one can associate a topological charge with the two Weyl nodes of opposing chirality. The only way to gap the isolated Weyl nodes is by merging them thereby annihilating their charges. On a more technical level, two dimensional representations of Dirac nodes, {\it i.e.}, Weyl nodes in 3D exhaust all Pauli matrices and therefore any mass term added can be compensated for by a shift of the momentum coordinate~\cite{burkov2011weyl}.   
In DSMs an even number of two dimensional representations of the Dirac theory (Weyl nodes) are degenerate in energy and momentum space and so no momentum transfer is required to annihilate them, meaning any coupling of the two nodes can gap them rendering them unstable. Concerning interactions, in principle the same physics as in graphene can be expected and also a symmetry broken gapped phase is possible with a mass term connecting the two chiral nodes. This has to be contrasted from WSMs which possess an intrinsic stability against many-body interactions. Weyl nodes are single two-dimensional representations of Dirac theories and their respective partners live in different locations in momentum space. This implies that a single Weyl cone cannot be gapped due to a simple mass term without a large momentum transfer. Long-range Coulomb interaction in principle is able to connect two Weyl cones thereby dynamically generating a mass. In three dimensions inelastic Coulomb scattering decays $\propto \frac{1}{k^2}$ in momentum space. If the Weyl nodes are separated by a distance $K$ in momentum space this suppresses internodal scattering, in perfect analogy to intervalley scattering in graphene. This implies that an even stronger Coulomb interaction than in graphene is required to generate a mass, which makes it very unlikely to occur. For that reason we refrain from studying the possibility of dynamic mass generation but instead focus on perturbative effects of Coulomb interaction assuming we do not connect/merge the two Weyl cones of opposing chirality. 

In order to study the effects of Coulomb interactions we use the perturbative renormalization group. It turns out that a modification of the more standard momentum shell scheme is more tailored towards the tilted cones: we integrate out excitations within an energy shell since there is no one-to-one correspondence between momentum and energy anymore once the cones are tilted~\cite{ShankarRevModPhys}. 
Our starting point is the following Hamiltonian
\begin{eqnarray}
  \label{eq: Hamiltonian WSM with tilting and Vc}
 {H} &=& \int d^3r \, \,\bar\psi_\alpha(\mathbf{r}) (-i\hbar v_F \mathbf{\nabla} \cdot \mathbf{\sigma} -i\hbar  v_F t \mathbf{d}\cdot \nabla) \psi_\alpha(\mathbf{r}) \nonumber \\ &+& \frac{e^2}{8\pi \epsilon} \int d^3 r \int d^3 r'\, \bar\psi_\alpha(\mathbf{r})\psi_\alpha(\mathbf{r}) \frac{1}{|\mathbf{r}-\mathbf{r}'|} \bar\psi_\beta(\mathbf{r}')\psi_\beta(\mathbf{r}')\nonumber \\
\end{eqnarray}
where double indices $\alpha,\beta=1,...,N$ are summed over and denote the number of Weyl nodes in the system (which we assume to have the same tilting). Furthermore, in writing this form of the Coulomb interaction we explicitly assume that there is no internodal scattering, justified by assuming the nodes are far away from each other in momentum space. 
In Eq.~\eqref{eq: Hamiltonian WSM with tilting and Vc}, $\epsilon = \epsilon_0 \epsilon_r$ is the product of the vacuum dielectric constant and the medium one, respectively. 
The interaction term introduced in the Hamiltonian describes an instantaneous interaction between the electric charges, thus we are neglecting any retardation effect that could come from the propagation of the photon field. The approximation is justified provided that $v_F/c \ll  1$, which will generically be the case (in graphene, for instance, $v_F/c\sim 1/300$).

The non-interacting Green function reads
\begin{eqnarray}
G_0(i\omega_n,\mathbf{k},\mathbf{d}) = \frac{\left(i\omega_n-v_F t\mathbf{d}\cdot\mathbf{k}\right) \sigma_0 + v_F\mathbf{k}\cdot \sigma}{(i\omega_n - v_Ft \mathbf{d}\cdot \mathbf{k})^2-v_F^2\mathbf{k}^2}\;.
\end{eqnarray}

We perform perturbation theory in the Coulomb interaction shown in Fig. \ref{fig: Coulomb interaction} with the vertex diagrams shown in Fig. \ref{fig: charge renormalization graphs} corresponding to its 1-loop order corrections.

\subsection{Self energy}

\begin{figure}
    \centering
        \begin{subfigure}[b]{0.2\textwidth}
        \includegraphics[width=\textwidth]{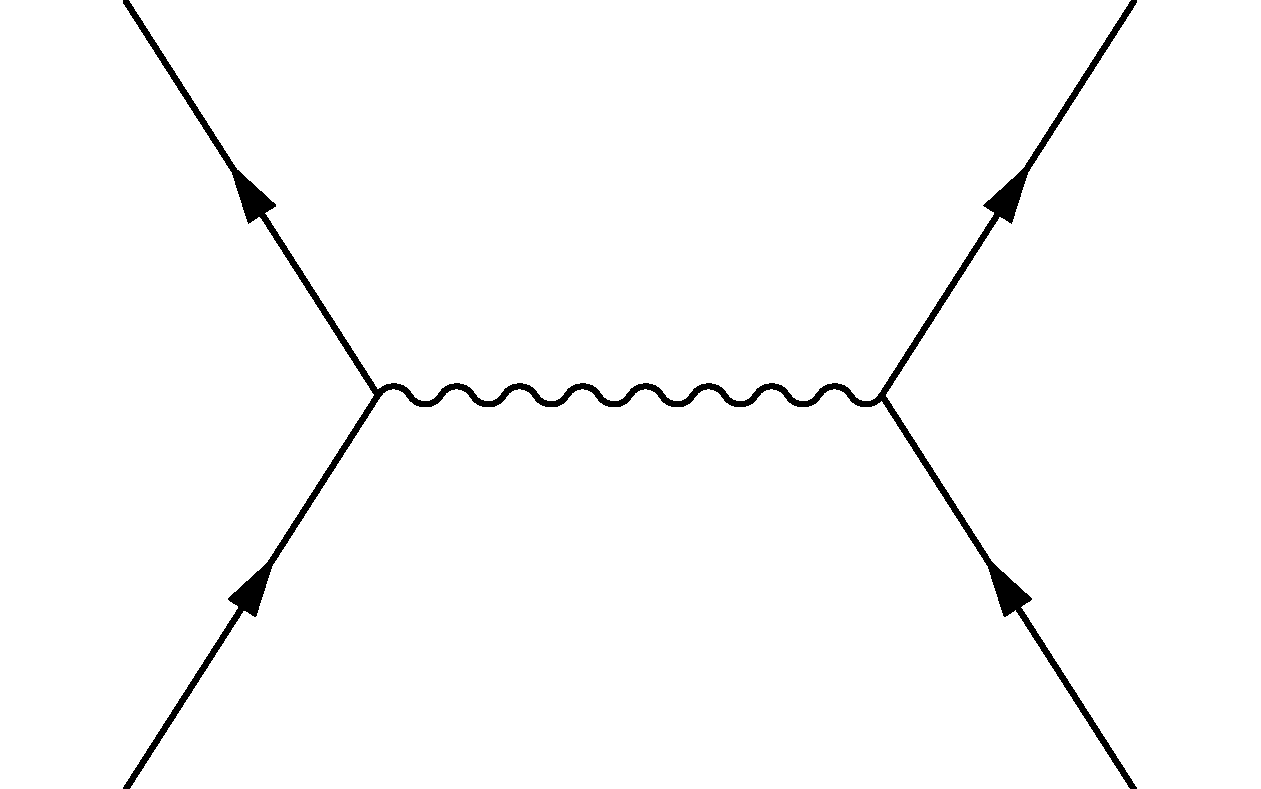}
  \caption{}
  \label{fig: Coulomb interaction}
  \end{subfigure}
  \qquad
    \begin{subfigure}[b]{0.2\textwidth}
        \includegraphics[width=\textwidth]{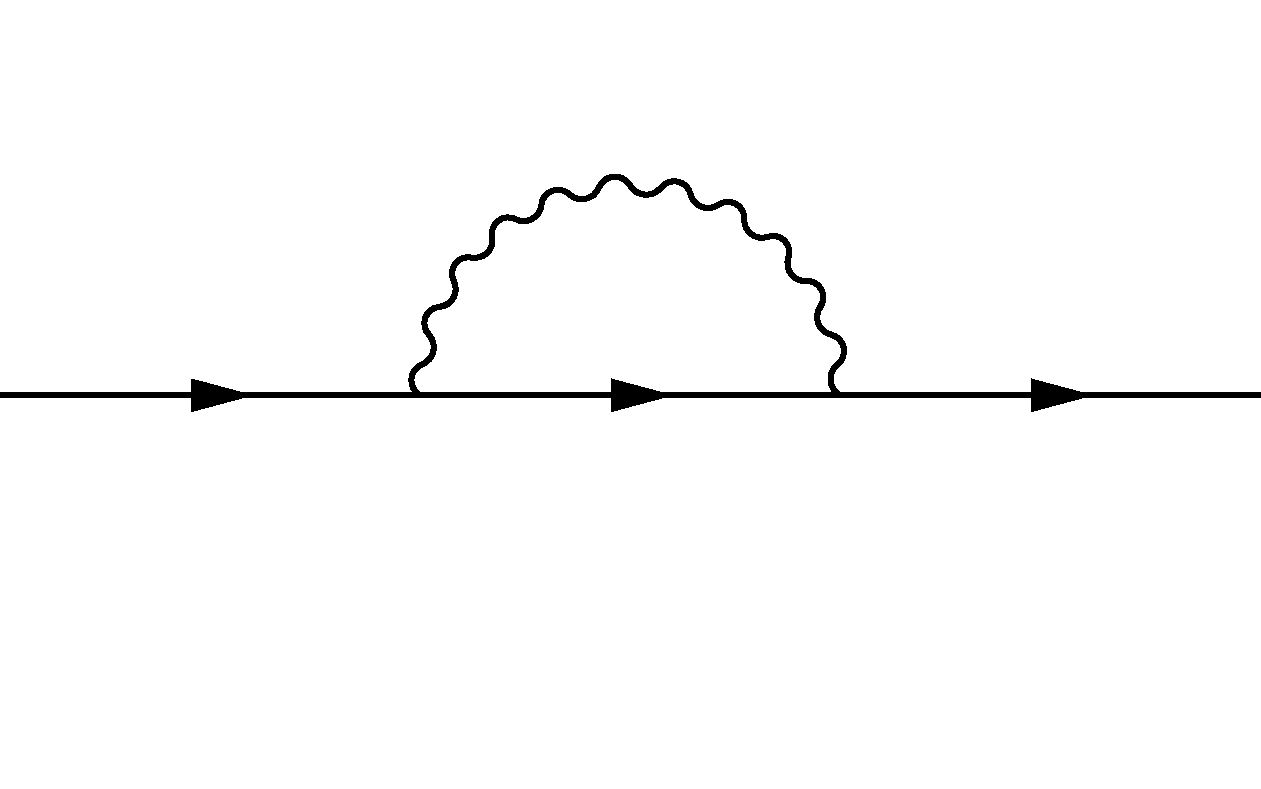}
  \caption{}
  \label{fig: self-energy graph}
    \end{subfigure}
    \caption{Coulomb vertex (a) and lowest order self-energy diagram (b)}
\end{figure}

To one-loop order in perturbation theory there is only one diagram contributing to the self-energy, shown in Fig. \ref{fig: self-energy graph}.

\begin{eqnarray}
  \label{eq: self-energy tilted WSM}
  \Sigma(i\omega,\mathbf{q}) = -T \sum_\nu \int \frac{d^3 k}{(2\pi)^3} \, G_0(i\omega_n +i\nu,\mathbf{k}+\mathbf{q}) V_c (-\mathbf{k})\nonumber \\
\end{eqnarray}
$V_c(-\mathbf{k}) = \frac{e^2}{\epsilon \mathbf{k}^2}$ being the Coulomb interaction and the Green's function is obtained from the Bloch Hamiltonian. In  the zero temperature limit the self-energy is given by
\begin{eqnarray}
  \Sigma(\omega,\mathbf{q}) &=&  \frac{e^2}{4 \epsilon v_F}  \int \frac{d^3 k}{(2\pi)^3}\, \frac{\mathbf{k}\cdot\sigma}{|\mathbf{k}||\mathbf{k}-\mathbf{q}|^2} \;.
\end{eqnarray}
Interestingly, it does not depend on the tilt $t$ and details of the derivation are shown in the Appendix~\ref{app:self}. Using a generalization of the Feynman trick
\begin{eqnarray}
\frac{1}{A^2|B|} = \frac{1}{2} \int_0^1 dx \, \frac{1}{\sqrt{1-x}} \frac{1}{[xA^2+(1-x)B^2]^{3/2}}\;
\end{eqnarray}
leads to
\begin{eqnarray}\label{eq:self}
   \Sigma(\omega,\mathbf{q}) &=& \frac{e^2}{8 \epsilon v_F}  \int_0^1 \frac{dx}{\sqrt{1-x}} \times \nonumber \\ &\times& \int \frac{d^3 k}{(2\pi)^3}\, \frac{(\mathbf{k}+(1-x)\mathbf{q})\cdot\sigma}{\left[\mathbf{k}^2+x(1-x)\mathbf{q}^2\right]^{3/2}}\nonumber \\ &=&\frac{e^2}{8 \epsilon v_F}  \int_0^1 \sqrt{(1-x)}dx \times \nonumber \\ &\times &\int \frac{d^3 k}{(2\pi)^3}\, \frac{\mathbf{q}\cdot\sigma}{\left[\mathbf{k}^2+x(1-x)\mathbf{q}^2\right]^{3/2}}\;.
\end{eqnarray}

We can integrate the momentum integral of the self energy within an energy window set by $\Lambda$ and $\Lambda'$. In the limit $v_F q \ll  \Lambda,\Lambda'$ Eq.~\eqref{eq:self} reads
\begin{eqnarray}\label{eq:selfscale}
 \Sigma(\omega,\mathbf{q}) =\frac{e^2}{12 \pi^2 \epsilon} \frac{\mathbf{q}\cdot\sigma}{1-t^2} \,\ln\left(\frac{\Lambda}{\Lambda'}\right)\;.
\end{eqnarray}

\subsection{Vertex corrections}

\begin{figure}
    \centering
    \begin{subfigure}[b]{0.2\textwidth}
        \includegraphics[width=\textwidth]{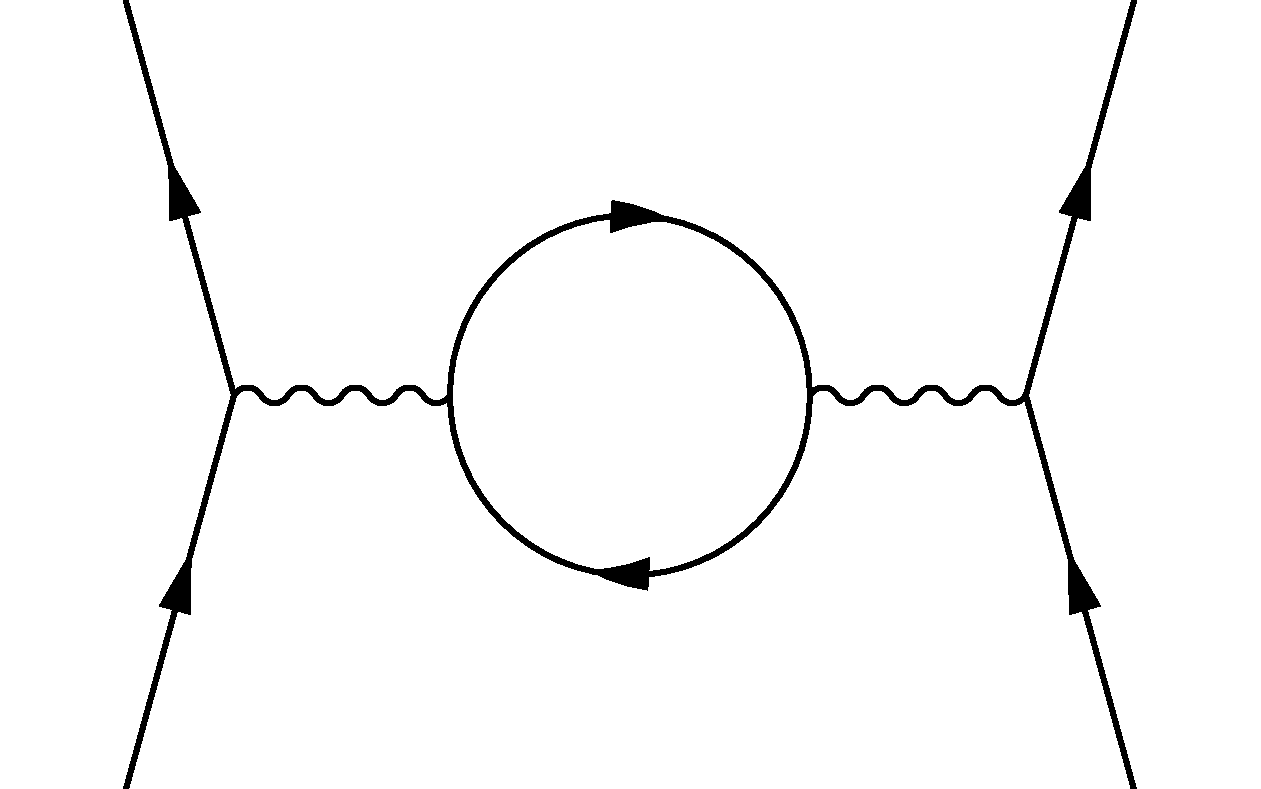}
        \caption{}
        \label{fig: charge renormalization a}
    \end{subfigure}
    \qquad
    \begin{subfigure}[b]{0.2\textwidth}
        \includegraphics[width=\textwidth]{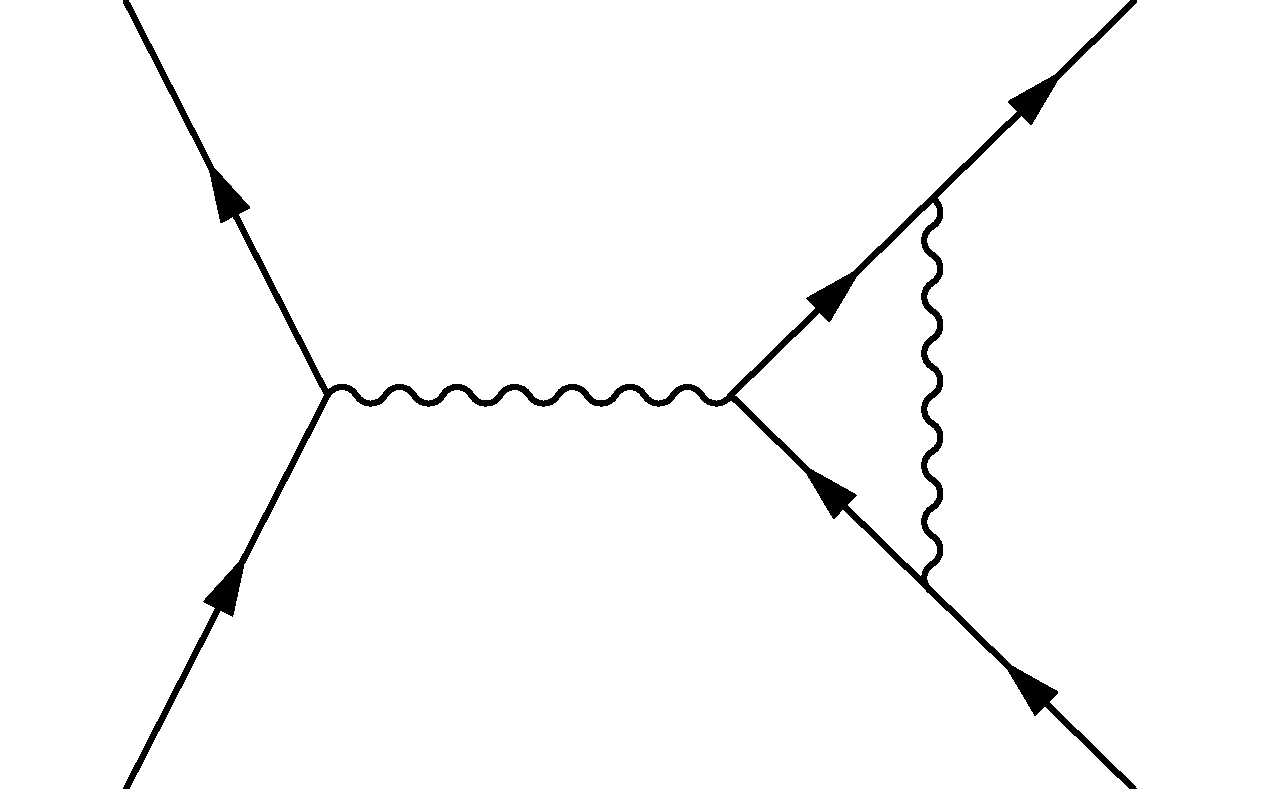}
        \caption{}
        \label{fig: charge renormalization b}
    \end{subfigure}
   \qquad
    \begin{subfigure}[b]{0.2\textwidth}
        \includegraphics[width=\textwidth]{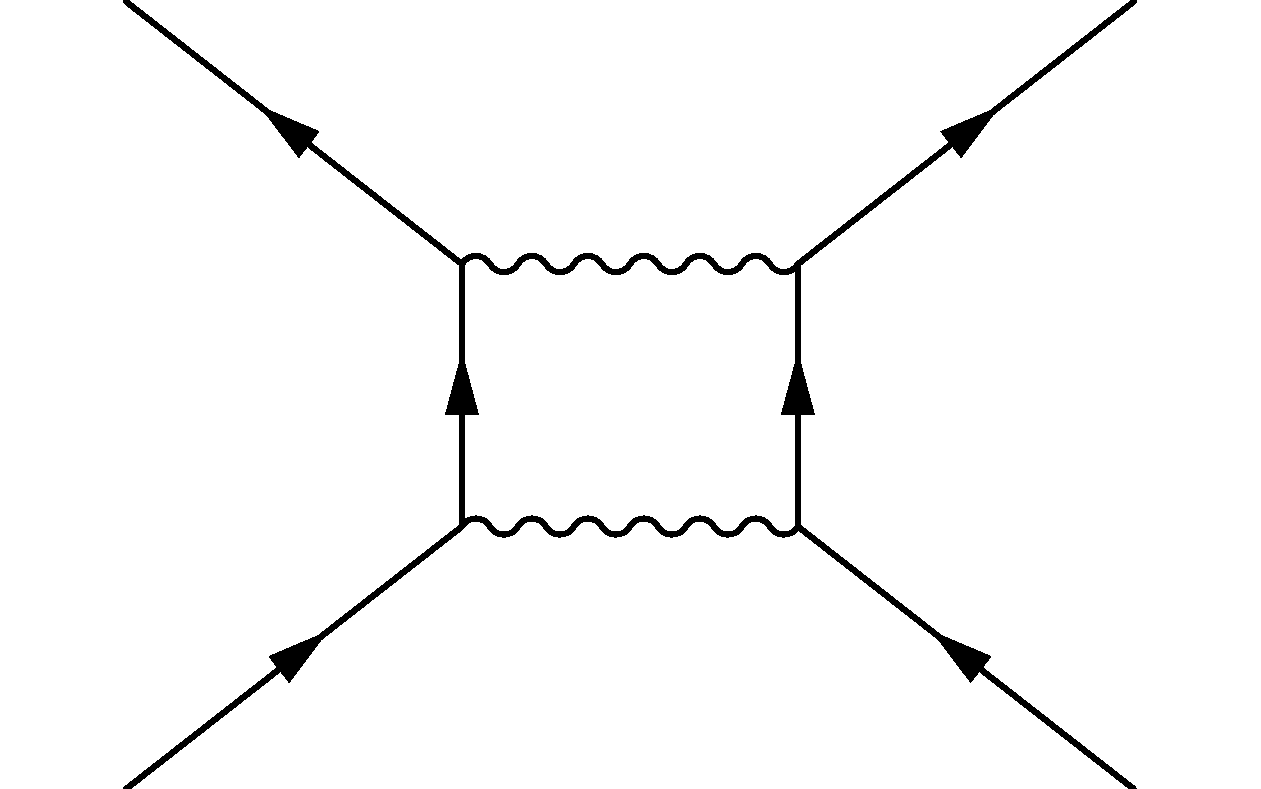}
        \caption{}
        \label{fig: charge renormalization c}
    \end{subfigure}
    \caption{Feynman graphs contributing to the charge renormalization up to one-loop corrections.}\label{fig: charge renormalization graphs}
\end{figure}

The Feynman graphs that contribute at one loop to the renormalization are shown in Fig. \ref{fig: charge renormalization graphs}.
\medskip\\
It turns out that only Fig.~\ref{fig: charge renormalization a} contributes logarithmically with the other two being zero to leading logarithmic order. It is given by
\begin{eqnarray}\label{eq:vertexscale}
I_a= N \left( \frac{e^2}{2\epsilon|\mathbf{q}|} \right)^2 \frac{1}{v_F} \frac{1}{6\pi^2} \frac{1}{1-t^2} \ln\left( \frac{\Lambda}{\Lambda'}\right)\;.
\end{eqnarray}
Details of the calculation are relegated to Appendix~\ref{app:vertex}.
\section{The flow equations}

From Eq.~\eqref{eq:selfscale} we can derive the renormalization of the  Fermi velocity $v_F$ given by
\begin{eqnarray}
\label{eq: rg eq fermi velocity}
  \frac{d\,v_F}{d\ln\Lambda} = v_F \frac{\alpha }{12\pi^2} \frac{1}{1-t^2}
\end{eqnarray}
where
\begin{eqnarray}
\alpha=\frac{e^2}{\epsilon v_F}\;
\end{eqnarray}
is the fine structure constant. This implies that, like in graphene, the Fermi velocity renormalizes towards higher values upon lowering the energy scale. 
An important observation here is that to one-loop order the combination $v_F t$ is an invariant under the renormalization group transformation, meaning $\frac{d\,v_F t}{d\ln\Lambda}=0$. This implies that $v_F \frac{d\,t}{d\ln\Lambda}=-t \frac{d\,v_F}{d\ln\Lambda}$ from which it follows that
\begin{eqnarray}
\label{eq:rgtilt}
  \frac{d\,t}{d\ln\Lambda} = -t \frac{\alpha}{12\pi^2} \frac{1}{1-t^2}\;.
\end{eqnarray}
An interpretation at this point is thus that the tilt renormalizes to zero upon lowering the energy scale, meaning asymptotically the theory becomes upright and isotropic again. 

From Eq.~\eqref{eq:vertexscale} we can derive the renomalization of the charge and dielectric constant (philosophically, only the dielectric constant is renormalized) which is given by
\begin{eqnarray}
\frac{d}{d\ln\Lambda} \left(\frac{e^2}{\epsilon}\right) =- N\frac{\alpha e^2}{24\pi^2 \epsilon} \frac{1}{1-t^2}\;.
\end{eqnarray}

Taking into account the flow of the Fermi velocity we can determine the flow of the fine structure constant $\alpha$ which turns out to be given by
\begin{eqnarray}\label{eq: rg eq coupling}
  \frac{d\alpha}{d\ln\Lambda} = -(N+2)\frac{\alpha^2}{24\pi^2} \frac{1}{1-t^2}
\end{eqnarray}

Eq.~\eqref{eq: rg eq fermi velocity}, Eq.~\eqref{eq:rgtilt}, and Eq.~\eqref{eq: rg eq coupling} provide the full set of flow equations. Again, these equations faithfully reproduce the renormalization flow in the untilted case for $t=0$~\cite{hosur2012charge}

In total we thus find that the fine structure constant flows to zero for low-energies, meaning asymptotically, the theory is non-interacting. Furthermore, the Fermi velocity $v_F$ renormalizes to higher values, while the tilt vanishes, thereby restoring isotropy of the theory. It is worthwhile noting that all of the above equations remain true in the presence of finite temperature and chemical potential as long as the running cutoff is bigger than either of the two. The flow will then stop at the larger of the two scales.

\section{Conclusion and Outlook}\label{sec:conc}

In this paper we studied the effect of a tilt term added to a Weyl node in observable quantities such as the optical conductivity and the polarization function. We found that in the extreme optical limit, {\it i.e.}, $\omega/T \gg  1$ the optical conductivity has no directional dependence, while for $\omega/T=\mathcal{O}(1)$ there is an observable effect. The polarization function and with that the plasmon spectrum shows some directional dependence which might be measurable in experiments. In a last part we studied the effect of long-range Coulomb interactions. Our main finding is that in the limit of low energies the theory becomes asymptotically isotropic again. 

{\it{Outlook}:} In a more recent work which is soon to appear one of the authors finds that in contrast to Coulomb interactions disorder in tilted Weyl systems increases the tilt. Therefore a very interesting interplay between interactions and disorder can be expected which is worthwhile investigating.

{\it{Acknowledgement:}}
This work is part of the D-ITP consortium, a program of the Netherlands Organisation for Scientific Research (NWO) that is funded by the Dutch Ministry of Education, Culture and Science (OCW). We acknowledge discussions and collaborations on related subjects with Tycho Sikkenk.

\appendix

\section{Optical conductivity for a massless Fermion field with tilting term}\label{app:optcon}

We consider a system described by the Hamiltonian given in Eq.~\eqref{eq:ham}. We can recast it in a more convenient form for the calculations to come,
\begin{eqnarray}
 H({\bf{k}},{\bf{d}}) &=& v_F
 \begin{pmatrix}
h_+ & 0 \\
 0 & h_-
\end{pmatrix} \; {\rm{with}}
\nonumber \\ 
h_{\pm}({\bf{k}},\mathbf{d}) &=& t {\bf{d}} \cdot {\bf{k}} \pm  {\bf{k}}\cdot\vec{\sigma}  \;.
\end{eqnarray}
The Green's function is defined as
\begin{eqnarray}
(i\omega_n - H(\mathbf{k},\mathbf{d})) G(i\omega_n,\mathbf{k},\mathbf{d}) = \mathbb{I}_4
\end{eqnarray}
leading to

\begin{eqnarray}
\label{eq: green fct tilted WSM}
G(i\omega_n,\mathbf{k},\mathbf{d}) = \frac{\big[ (i\omega_n - t v_F \mathbf{d}\cdot \mathbf{k}) \otimes \mathbb{I}_4 +  v_F\tau_z \otimes (\mathbf{k}\cdot \mathbf{\sigma}) \big]}{(i\omega_n - t v_F\mathbf{d}\cdot \mathbf{k})^2- v_F^2\mathbf{k}^2} \nonumber \\
\end{eqnarray}
The current operators are computed in the standard fashion
\begin{eqnarray}
j^{\mu} = e \frac{\partial H}{\partial k_\mu}
\quad
\longrightarrow
\quad
j^i=ev_F(  \tau_z \otimes \sigma^i + t d_i)
\end{eqnarray}
We introduce the spectral function 
\begin{eqnarray}
 && A(\omega^+,\mathbf{q},\mathbf{d}) = \frac{\pi}{v_F  |\mathbf{q}|}\big[ (\omega - t v_F\mathbf{d}\cdot \mathbf{q}) +  v_F\tau_z \otimes (\mathbf{q}\cdot \mathbf{\sigma}) \big]\times \nonumber \\ && \Big[ \delta \big(\omega -v_F t\mathbf{d}\cdot\mathbf{q}- v_F|\mathbf{q}| \big) -\delta \big(\omega -v_Ft\mathbf{d}\cdot\mathbf{q} +  v_F|\mathbf{q}| \big) \Big]\nonumber \\
\end{eqnarray}
and consider processes with momentum transfer ${\bf{k}}=\bf{0}$
\begin{widetext}
\begin{eqnarray}
K^{ij}(i\omega_n) =-\frac{e^2v_F^2}{\beta} \sum_m \int \frac{d^3 q}{(2\pi)^3} \int \frac{d\omega'}{2\pi} \int \frac{d\omega''}{2\pi} \, \mathrm{Tr} \bigg[ (  \tau_z \otimes \sigma^i + t d_i) \frac{A(\omega',\mathbf{q})}{i\omega_n+i\omega_m-\omega'}  (  \tau_z \otimes \sigma^j + t d_j) \frac{A(\omega'',\mathbf{q})}{i\omega_m-\omega''} \bigg]\;.
\end{eqnarray}

We first sum over Matsubara frequencies, then perform analytic continuation to real frequency and finally integrate over $\omega''$

\begin{eqnarray}
\mathrm{Im} K^{ij}(\omega^+) =-e^2 v_F^2 \int \frac{d^3 q}{(2\pi)^3} \int \frac{d\omega'}{4\pi} \,\big[n_F(\omega')-n_F(\omega'-\omega) \big] 
\mathrm{Tr} \bigg[ (  \tau_z \otimes \sigma^i + td_i) {A(\omega',\mathbf{q})} (  \tau_z \otimes \sigma^j + td_j) {A(\omega'-\omega,\mathbf{q})} \bigg] \nonumber \\
\end{eqnarray}
yielding
\begin{eqnarray}
\mathrm{Im}K^{ij}(\omega^+,\mathbf{0}) =-\frac{e^2}{v_F} \sum_{\alpha=\pm}\,\int \frac{d^3 q}{32\pi^2} \Bigg\{ \frac{n_F(t\mathbf{d}\cdot\mathbf{q}+\alpha  |\mathbf{q}|)-n_F(t\mathbf{d}\cdot\mathbf{q}+\alpha  |\mathbf{q}|-\omega)}{\mathbf{q}^2}
\Big[ \delta \big(-\omega \big) -\delta \big(2\alpha  |\mathbf{q}|- \omega  \big) \Big] \times \\ \nonumber \times 
\mathrm{Tr} \bigg[ (  \tau_z \otimes \sigma^i + td_i) (\alpha  |\mathbf{q}|+  \tau_z \otimes(\mathbf{q}\cdot\mathbf{\sigma})) (  \tau_z \otimes \sigma^j + td_j) (\alpha  |\mathbf{q}|-\omega+  \tau_z \otimes(\mathbf{q}\cdot\mathbf{\sigma})) \bigg] 
\Bigg\}
\end{eqnarray}

where we introduced $\alpha=\pm 1$ and rescaled the momentum $v_F{\bf{q}} \to  {\bf{q}}$. We may drop the term $\delta(-\omega)$ since the integrand function is zero for $\omega=0$.

\begin{eqnarray}
\mathrm{Im}K^{ij}(\omega^+,\mathbf{0}) =\frac{e^2}{8\pi^2v_F} \sum_{\alpha=\pm} \int {d^3 q} \:  T^{ij} \: \Bigg\{ \frac{n_F(t\mathbf{d}\cdot\mathbf{q}+\alpha  |\mathbf{q}|)-n_F(t\mathbf{d}\cdot\mathbf{q}+\alpha  |\mathbf{q}|-\omega)}{\mathbf{q}^2}  \delta \left(2\alpha  |\mathbf{q}|- \omega  \right)  \Bigg\}
\end{eqnarray}
\end{widetext}
Without loss of generality we take the direction of the tilting to be $\mathbf{d}=(0,0,1)$ and look at the conductivities in the directions parallel and perpendicular to $\mathbf{d}$ to see whether it carries a signature of the tilting term.\\
In particular we define $\sigma_{\parallel}=\sigma^{zz}$ and $\sigma_{\perp}=(\sigma^{xx}+\sigma^{yy})/2$.\\
Switching to spherical coordinates, $(q_x,q_y,q_z) \rightarrow (q\sin\theta\cos\phi, q\sin\theta\sin\phi, q\cos\theta)$, we find
\begin{eqnarray}
\frac{T^{xx}+T^{yy}}{2} = -\alpha q \omega + q^2 \sin^2\theta
\end{eqnarray} and
\begin{eqnarray}
T^{zz} = -\alpha q\omega +t^2   q(2\alpha   q-\omega) + 2 q^2 \cos^2\theta + t (2\alpha   q -\omega) q\cos\theta \;.\nonumber \\
\end{eqnarray}
Using elementary manipulations we find
\begin{widetext}
\begin{eqnarray}
\mathrm{Im} \bigg[\frac{K^{xx}+K^{yy}}{2} \bigg] 
&=&\frac{e^2}{8\pi^2v_F}\sum_{\alpha=\pm}\,\int_{0}^{2\pi}d\phi \int_{0}^{\pi} d\theta \int_{0}^{\infty} dq \, q^2 \sin\theta \frac{n_F(tq\cos\theta+\alpha  {q})-n_F(tq\cos\theta+\alpha  q-\omega)}{ {q}^2} \times \nonumber \\ 
&& \times 
\delta \big(2\alpha  {q}- \omega  \big)
\left( -\alpha q \omega + q^2 \sin^2\theta \right) \nonumber \\
&=& \frac{e^2}{4\pi v_F} \sum_{\alpha=\pm}\, \int_{-1}^{1}dx \int_{0}^{\infty} dq \, \bigg[ {n_F(tqx+\alpha  {q})-n_F(tqx+ \alpha  q-\omega)} \bigg] 
\frac{1}{2 }\delta \left({q}- \frac{\omega}{2\alpha  }  \right) 
\left( -\alpha q \omega + q^2 (1-x^2) \right) \nonumber \\
&=& -\frac{e^2\omega^2}{32\pi v_F } \int_{-1}^{1} dx \, \bigg[ n_F \left(\frac{\omega}{2}(tx+1) \right)-n_F\left( \frac{\omega}{2}(tx-1)\right) \bigg] (1+x^2) \;.
\end{eqnarray}
\end{widetext}
Notice that the terms with $\alpha = \pm1$ correspond to positive and negative frequencies, respectively.\\
This directly leads to the perpendicular conductivity given in Eq.~\eqref{eq:sigmaperp}.
 In an analoguous way we compute the parallel conductivity. For $\omega$>0:
\begin{widetext}
\begin{eqnarray}
\mathrm{Im} \bigg[K^{zz} \bigg] 
&=&\frac{e^2}{8\pi^2 v_F}\sum_{\alpha=\pm}\,\int_{0}^{2\pi}d\phi \int_{0}^{\pi} d\theta \int_{0}^{\infty} dq \, q^2 \sin\theta \frac{n_F(tq\cos\theta+\alpha  {q})-n_F(tq\cos\theta+\alpha  q-\omega)}{ {q}^2} \times \nonumber \\ 
&&  \times 
 \delta \big(2 {q}- \omega  \big)
\left( -\alpha q\omega +t^2   q(2\alpha   q-\omega) + 2 q^2 \cos^2\theta + t (2\alpha   q -\omega) q\cos\theta \right) \nonumber \\
&=& \frac{e^2}{4\pi v_F } \sum_{\alpha=\pm}\, \int_{-1}^{1} dx \int_{0}^{\infty} dq \, \bigg[ {n_F(tqx+\alpha  {q})-n_F(tqx+\alpha  q-\omega)} \bigg] \times \nonumber \\ 
& &  \times 
\frac{1}{2 }\delta \left({q}- \frac{\omega}{2\alpha  }  \right)
\left( -\alpha q\omega +t^2   q(2\alpha   q-\omega) + 2 q^2 x^2 + t (2\alpha   q -\omega) qx \right) \nonumber \\
&=& -\frac{e^2\omega^2}{16\pi v_F } \int_{-1}^{1} dx \, \bigg[ {n_F \left( \frac{\omega}{2}(tx+1)\right)-n_F\left(\frac{\omega}{2}(tx-1)\right)} \bigg] (1-x^2) 
\end{eqnarray}
\end{widetext}
which leads to Eq.~\eqref{eq:sigmaparallel}.

The effect of the tilting of the Dirac cones is a rather indirect contribution contained in the Fermi-Dirac distribution.
A simple check of the result can be made by setting $t=0$. We compute the longitudinal conductivity:
\begin{eqnarray}
  \sigma_L(\omega) &=& \sum_i\frac{\sigma^{ii}}{3} = \frac{2 \sigma_{\perp}}{3} + \frac{\sigma_{\parallel}}{3}  \nonumber \\
  &=& -\frac{e^2\omega}{24\pi v_F }  \int_{-1}^{1} dx \, \bigg[ {n_F\left(\frac{\omega}{2}\right)-n_F\left(-\frac{\omega}{2}\right)} \bigg] \nonumber \\ &=& -\frac{e^2\omega}{12\pi v_F}  \bigg[ {n_F\left(\frac{\omega}{2}\right)-n_F\left(-\frac{\omega}{2}\right)} \bigg]
\end{eqnarray}
which is precisely the result obtained for free massless fermions. It also retains the correct behavior in the zero-temperature limit, 
\begin{equation}
\label{eq: ac conductivity DSM}
   \sigma_L(\omega) \stackrel{T\rightarrow 0}{=} \frac{e^2|\omega|}{12\pi v_F}
\end{equation}
in agreement with previous work, \cite{hosur2012charge}.

\section{Polarization function}\label{app:polfun}

\begin{widetext}
We evaluate the polarizability from
\begin{eqnarray}
  \pi (i\nu,\mathbf{q}/v_F,\mathbf{d}) &=-\frac{2}{\beta v_F^3} \sum_m \int \frac{d^3 k}{(2\pi)^3} \frac{(i\omega_m-t \mathbf{d}\cdot\mathbf{k})(i\nu+i\omega_m-t\mathbf{d}\cdot(\mathbf{k}+\mathbf{q}))+\mathbf{k}\cdot(\mathbf{k}+\mathbf{q})}{\big[(i\omega_m -t\mathbf{d}\cdot\mathbf{k})^2 - \mathbf{k}^2\big] \big[(i\omega_m +i\nu -t\mathbf{d}\cdot(\mathbf{k}+\mathbf{q})) ^2 - |\mathbf{k}+\mathbf{q}|^2 \big]} 
\end{eqnarray}
and using standard manipulations we canbring it into the form
\begin{equation}
  \pi(i\nu,\mathbf{q}/v_F,\mathbf{d}) = -\int \frac{d^3k}{(2\pi)^3v_F^3} \sum_{s,s'=\pm 1} \frac{n_F(s|\mathbf{k}|+t\mathbf{d}\cdot\mathbf{k})-n_F(s'|\mathbf{k}+\mathbf{q}|+t\mathbf{d}\cdot(\mathbf{k}+\mathbf{q}))}{i\nu-t\mathbf{d}\cdot\mathbf{q}+s|\mathbf{k}|-s'|\mathbf{k}+\mathbf{q}|} \times \frac{1+ss'\cos\Theta}{2}\;,
\end{equation}
where $\Theta$ is the angle between vectors $\mathbf{k}$ and $\mathbf{k}+\mathbf{q}$.
We notice that at zero temperature due to $t<1$ none of the occupation
numbers are changed by the tilt, so we can set the tilt $t=0$ in
the nominator. We can also switch the external frequency $i\nu-t\mathbf{d}\cdot\mathbf{q}\rightarrow i\nu$
to get rid of the tilt inside the integrand completely. What remains
is the untilted Weyl fermion polarization function at zero chemical potential and zero
temperature, which can be calculated in a standard way. Then we switch
the variables back to finally obtain the retarded polarization function, 
\begin{eqnarray}\label{eq:polfuncim}
  \pi^R(\nu,\mathbf{q}) &=& \frac{q^2}{24\pi^2 v_F} \left\{ \ln\left(1+\frac{4\Lambda^2}{v_F^2 q^2-(\nu-tv_F\mathbf{d}\cdot\mathbf{q})^2}\right) + \right. \nonumber \\
  &+& \left. \frac{8\Lambda^3}{(v_F^2 q^2-(\nu-t v_F  \mathbf{d}\cdot\mathbf{q})^2)^{3/2}} \arctan\left(\frac{\sqrt{v_F^2 q^2-(\nu-t v_F \mathbf{d}\cdot\mathbf{q})^2}}{2\Lambda}\right) - 
\frac{4\Lambda^2}{v_F^2 q^2-(\nu-tv_F \mathbf{d}\cdot\mathbf{q})^2} \right\}
\end{eqnarray}
where the remaining steps leading to Eq.~\eqref{eq:pol} are elementary and thus not shown explicitly.
\end{widetext}

\section{Frequency integral}\label{app:self}
Since Coulomb interaction does not depend on momentum one can absorb the external momentum in the frequency integral by a simple shift in the integration variable. Therefore, the integral reads
\begin{eqnarray}
\int_{-\infty}^{\infty} \frac{d\nu}{2\pi}\, \frac{i\nu -t v_F\mathbf{d}\cdot\mathbf{k} +v_F \mathbf{k}\cdot \sigma}{(i\nu - t v_F \mathbf{d}\cdot \mathbf{k})^2-v_F^2|\mathbf{k}|^2} \;.
\end{eqnarray}

Since the integral formally is divergent we introduce a cut-off $\Lambda$ that will be removed later:
\begin{eqnarray}
&&\int_{-\Lambda}^{\Lambda} \frac{d\nu}{2\pi}\, \frac{i\nu -t v_F\mathbf{d}\cdot\mathbf{k} +v_F \mathbf{k}\cdot \sigma}{(i\nu - t v_F \mathbf{d}\cdot \mathbf{k})^2-v_F^2|\mathbf{k}|^2}\nonumber \\ &=& 
\frac{1}{2\pi} \mathrm{arctan} \left( \frac{2 \Lambda t v_F \mathbf{d}\cdot\mathbf{k}}{\Lambda^2+v_F^2|\mathbf{k}|^2-t^2 v_F^2 (\mathbf{d}\cdot\mathbf{k})^2} \right) \nonumber \\ &-& \frac{\mathbf{k}\cdot\sigma}{2\pi|\mathbf{k}|}  \left[ \mathrm{arctan} \left( \frac{\Lambda}{v_F|\mathbf{k}|+tv_F\mathbf{d}\cdot\mathbf{k}} \right) \right. \nonumber \\  &+& \left. \mathrm{arctan} \left( \frac{\Lambda}{v_F|\mathbf{k}|-tv_F\mathbf{d}\cdot\mathbf{k}} \right) \right]\;.
\end{eqnarray}
Taking the limit for $\Lambda\rightarrow\infty$ we find the result

\begin{equation}
\lim_{\Lambda \to \infty}  \int_{-\Lambda}^{\Lambda} \frac{d\nu}{2\pi}\, \frac{i\nu -tv_F\mathbf{d}\cdot\mathbf{k} + v_F\mathbf{k}\cdot \sigma}{(i\nu - tv_F \mathbf{d}\cdot \mathbf{k})^2-v_F^2|\mathbf{k}|^2}
  = - \frac{\mathbf{k}\cdot\sigma}{2|\mathbf{k}|} \;.
\end{equation}
An alternative derivation can be performed using residues.

\section{Vertex corrections}\label{app:vertex}
It is relatively straightforward to relate the diagram shown in Fig.~\ref{fig: charge renormalization a} to the polarization function calculated in Eq.~\eqref{eq:polfuncim} since it contains the polarization bubble. Following the discussion of Appendix B, the tilt $t$ drops out of
the integrand, however the integration boundaries are set by an energy
cutoff, which is still tilt-dependent. In this way, the tilt parameter
$t$ enters the final expression for diagrams as a prefactor. The
result up to logarithmic accuracy is then given by

\begin{eqnarray}
I_a(i\nu,\mathbf{q}) &=& N \left( \frac{e^2}{2\epsilon|\mathbf{q}|^2} \right)^2 \frac{1}{1-t^2} \pi(\nu,{\bf{q}})  \nonumber \\ &=& N \left( \frac{e^2}{2\epsilon|\mathbf{q}|^2} \right)^2 \frac{1}{6\pi^2}\frac{1}{1-t^2} \ln \left( \frac{\Lambda}{\Lambda'}\right)\;.
\end{eqnarray}

The same considerations hold for the vertex diagram 3-b. After the Matsubara summation, the tilt is eliminated from the integrand and then
following the discussion of Eq.~(35) in Ref.~[\onlinecite{PhysRevB.89.235431}] the vertex function is zero.


\begin{thebibliography}{29}%
\makeatletter
\providecommand \@ifxundefined [1]{%
 \@ifx{#1\undefined}
}%
\providecommand \@ifnum [1]{%
 \ifnum #1\expandafter \@firstoftwo
 \else \expandafter \@secondoftwo
 \fi
}%
\providecommand \@ifx [1]{%
 \ifx #1\expandafter \@firstoftwo
 \else \expandafter \@secondoftwo
 \fi
}%
\providecommand \natexlab [1]{#1}%
\providecommand \enquote  [1]{``#1''}%
\providecommand \bibnamefont  [1]{#1}%
\providecommand \bibfnamefont [1]{#1}%
\providecommand \citenamefont [1]{#1}%
\providecommand \href@noop [0]{\@secondoftwo}%
\providecommand \href [0]{\begingroup \@sanitize@url \@href}%
\providecommand \@href[1]{\@@startlink{#1}\@@href}%
\providecommand \@@href[1]{\endgroup#1\@@endlink}%
\providecommand \@sanitize@url [0]{\catcode `\\12\catcode `\$12\catcode
  `\&12\catcode `\#12\catcode `\^12\catcode `\_12\catcode `\%12\relax}%
\providecommand \@@startlink[1]{}%
\providecommand \@@endlink[0]{}%
\providecommand \url  [0]{\begingroup\@sanitize@url \@url }%
\providecommand \@url [1]{\endgroup\@href {#1}{\urlprefix }}%
\providecommand \urlprefix  [0]{URL }%
\providecommand \Eprint [0]{\href }%
\providecommand \doibase [0]{http://dx.doi.org/}%
\providecommand \selectlanguage [0]{\@gobble}%
\providecommand \bibinfo  [0]{\@secondoftwo}%
\providecommand \bibfield  [0]{\@secondoftwo}%
\providecommand \translation [1]{[#1]}%
\providecommand \BibitemOpen [0]{}%
\providecommand \bibitemStop [0]{}%
\providecommand \bibitemNoStop [0]{.\EOS\space}%
\providecommand \EOS [0]{\spacefactor3000\relax}%
\providecommand \BibitemShut  [1]{\csname bibitem#1\endcsname}%
\let\auto@bib@innerbib\@empty
%</preamble>
\bibitem [{\citenamefont {Wan}\ \emph {et~al.}(2011)\citenamefont {Wan},
  \citenamefont {Turner}, \citenamefont {Vishwanath},\ and\ \citenamefont
  {Savrasov}}]{wan2011topological}%
  \BibitemOpen
  \bibfield  {author} {\bibinfo {author} {\bibfnamefont {X.}~\bibnamefont
  {Wan}}, \bibinfo {author} {\bibfnamefont {A.~M.}\ \bibnamefont {Turner}},
  \bibinfo {author} {\bibfnamefont {A.}~\bibnamefont {Vishwanath}}, \ and\
  \bibinfo {author} {\bibfnamefont {S.~Y.}\ \bibnamefont {Savrasov}},\
  }\href@noop {} {\bibfield  {journal} {\bibinfo  {journal} {Phys. Rev. B}\
  }\textbf {\bibinfo {volume} {83}},\ \bibinfo {pages} {205101} (\bibinfo
  {year} {2011})}\BibitemShut {NoStop}%
\bibitem [{\citenamefont {Balents}(2011)}]{balents2011viewpoint}%
  \BibitemOpen
  \bibfield  {author} {\bibinfo {author} {\bibfnamefont {L.}~\bibnamefont
  {Balents}},\ }\href@noop {} {\bibfield  {journal} {\bibinfo  {journal}
  {Physics}\ \textbf {\bibinfo {volume} {4}} ,\ \bibinfo {pages} {36}} (\bibinfo {year} {2011})}\BibitemShut
  {NoStop}%
\bibitem [{\citenamefont {Weng}\ \emph {et~al.}(2015)\citenamefont {Weng},
  \citenamefont {Fang}, \citenamefont {Fang}, \citenamefont {Bernevig},\ and\
  \citenamefont {Dai}}]{weng2015weyl}%
  \BibitemOpen
  \bibfield  {author} {\bibinfo {author} {\bibfnamefont {H.}~\bibnamefont
  {Weng}}, \bibinfo {author} {\bibfnamefont {C.}~\bibnamefont {Fang}}, \bibinfo
  {author} {\bibfnamefont {Z.}~\bibnamefont {Fang}}, \bibinfo {author}
  {\bibfnamefont {B.~A.}\ \bibnamefont {Bernevig}}, \ and\ \bibinfo {author}
  {\bibfnamefont {X.}~\bibnamefont {Dai}},\ }\href@noop {} {\bibfield
  {journal} {\bibinfo  {journal} {Phys. Rev. X}\ }\textbf {\bibinfo {volume}
  {5}},\ \bibinfo {pages} {011029} (\bibinfo {year} {2015})}\BibitemShut
  {NoStop}%
\bibitem [{\citenamefont {Huang}\ \emph {et~al.}(2015)\citenamefont {Huang},
  \citenamefont {Xu}, \citenamefont {Belopolski}, \citenamefont {Lee},
  \citenamefont {Chang}, \citenamefont {Wang}, \citenamefont {Alidoust},
  \citenamefont {Bian}, \citenamefont {Neupane}, \citenamefont {Zhang} \emph
  {et~al.}}]{huang2015weyl}%
  \BibitemOpen
  \bibfield  {author} {\bibinfo {author} {\bibfnamefont {S.-M.}\ \bibnamefont
  {Huang}}, \bibinfo {author} {\bibfnamefont {S.-Y.}\ \bibnamefont {Xu}},
  \bibinfo {author} {\bibfnamefont {I.}~\bibnamefont {Belopolski}}, \bibinfo
  {author} {\bibfnamefont {C.-C.}\ \bibnamefont {Lee}}, \bibinfo {author}
  {\bibfnamefont {G.}~\bibnamefont {Chang}}, \bibinfo {author} {\bibfnamefont
  {B.}~\bibnamefont {Wang}}, \bibinfo {author} {\bibfnamefont {N.}~\bibnamefont
  {Alidoust}}, \bibinfo {author} {\bibfnamefont {G.}~\bibnamefont {Bian}},
  \bibinfo {author} {\bibfnamefont {M.}~\bibnamefont {Neupane}}, \bibinfo
  {author} {\bibfnamefont {C.}~\bibnamefont {Zhang}},  \emph {et~al.},\
  }\href@noop {} {\bibfield  {journal} {\bibinfo  {journal} {Nature Comm.}\
  }\textbf {\bibinfo {volume} {6}},\ \bibinfo {pages} {7373} (\bibinfo {year}
  {2015})}\BibitemShut {NoStop}%
\bibitem [{\citenamefont {Turner}\ and\ \citenamefont
  {Vishwanath}(2013)}]{turner2013beyond}%
  \BibitemOpen
  \bibfield  {author} {\bibinfo {author} {\bibfnamefont {A.~M.}\ \bibnamefont
  {Turner}}\ and\ \bibinfo {author} {\bibfnamefont {A.}~\bibnamefont
  {Vishwanath}},\ }\href@noop {} {\bibfield  {journal} {\bibinfo  {journal}
  {Topological Insulators}\ }\textbf {\bibinfo {volume} {6}},\ \bibinfo {pages}
  {293} (\bibinfo {year} {2013})}\BibitemShut {NoStop}%
\bibitem [{\citenamefont {Lv}\ \emph {et~al.}(2015{\natexlab{a}})\citenamefont
  {Lv}, \citenamefont {Weng}, \citenamefont {Fu}, \citenamefont {Wang},
  \citenamefont {Miao}, \citenamefont {Ma}, \citenamefont {Richard},
  \citenamefont {Huang}, \citenamefont {Zhao}, \citenamefont {Chen},
  \citenamefont {Fang}, \citenamefont {Dai}, \citenamefont {Qian},\ and\
  \citenamefont {Ding}}]{Lv2015Experimental}%
  \BibitemOpen
  \bibfield  {author} {\bibinfo {author} {\bibfnamefont {B.~Q.}\ \bibnamefont
  {Lv}}, \bibinfo {author} {\bibfnamefont {H.~M.}\ \bibnamefont {Weng}},
  \bibinfo {author} {\bibfnamefont {B.~B.}\ \bibnamefont {Fu}}, \bibinfo
  {author} {\bibfnamefont {X.~P.}\ \bibnamefont {Wang}}, \bibinfo {author}
  {\bibfnamefont {H.}~\bibnamefont {Miao}}, \bibinfo {author} {\bibfnamefont
  {J.}~\bibnamefont {Ma}}, \bibinfo {author} {\bibfnamefont {P.}~\bibnamefont
  {Richard}}, \bibinfo {author} {\bibfnamefont {X.~C.}\ \bibnamefont {Huang}},
  \bibinfo {author} {\bibfnamefont {L.~X.}\ \bibnamefont {Zhao}}, \bibinfo
  {author} {\bibfnamefont {G.~F.}\ \bibnamefont {Chen}}, \bibinfo {author}
  {\bibfnamefont {Z.}~\bibnamefont {Fang}}, \bibinfo {author} {\bibfnamefont
  {X.}~\bibnamefont {Dai}}, \bibinfo {author} {\bibfnamefont {T.}~\bibnamefont
  {Qian}}, \ and\ \bibinfo {author} {\bibfnamefont {H.}~\bibnamefont {Ding}},\
  }\href@noop {} {\bibfield  {journal} {\bibinfo  {journal} {Phys. Rev. X}\
  }\textbf {\bibinfo {volume} {5}},\ \bibinfo {pages} {031013} (\bibinfo {year}
  {2015}{\natexlab{a}})}\BibitemShut {NoStop}%
\bibitem [{\citenamefont {Xu}\ \emph {et~al.}(2015{\natexlab{a}})\citenamefont
  {Xu}, \citenamefont {Belopolski}, \citenamefont {Alidoust}, \citenamefont
  {Neupane}, \citenamefont {Bian}, \citenamefont {Zhang}, \citenamefont
  {Sankar}, \citenamefont {Chang}, \citenamefont {Yuan}, \citenamefont {Lee}
  \emph {et~al.}}]{xu2015discovery}%
  \BibitemOpen
  \bibfield  {author} {\bibinfo {author} {\bibfnamefont {S.-Y.}\ \bibnamefont
  {Xu}}, \bibinfo {author} {\bibfnamefont {I.}~\bibnamefont {Belopolski}},
  \bibinfo {author} {\bibfnamefont {N.}~\bibnamefont {Alidoust}}, \bibinfo
  {author} {\bibfnamefont {M.}~\bibnamefont {Neupane}}, \bibinfo {author}
  {\bibfnamefont {G.}~\bibnamefont {Bian}}, \bibinfo {author} {\bibfnamefont
  {C.}~\bibnamefont {Zhang}}, \bibinfo {author} {\bibfnamefont
  {R.}~\bibnamefont {Sankar}}, \bibinfo {author} {\bibfnamefont
  {G.}~\bibnamefont {Chang}}, \bibinfo {author} {\bibfnamefont
  {Z.}~\bibnamefont {Yuan}}, \bibinfo {author} {\bibfnamefont {C.-C.}\
  \bibnamefont {Lee}},  \emph {et~al.},\ }\href@noop {} {\bibfield  {journal}
  {\bibinfo  {journal} {Science}\ }\textbf {\bibinfo {volume} {349}},\ \bibinfo
  {pages} {613} (\bibinfo {year} {2015}{\natexlab{a}})}\BibitemShut {NoStop}%
\bibitem [{\citenamefont {Xu}\ \emph {et~al.}(2015{\natexlab{b}})\citenamefont
  {Xu}, \citenamefont {Alidoust}, \citenamefont {Belopolski}, \citenamefont
  {Yuan}, \citenamefont {Bian}, \citenamefont {Chang}, \citenamefont {Zheng},
  \citenamefont {Strocov}, \citenamefont {Sanchez}, \citenamefont {Chang} \emph
  {et~al.}}]{xu2015discovery2}%
  \BibitemOpen
  \bibfield  {author} {\bibinfo {author} {\bibfnamefont {S.-Y.}\ \bibnamefont
  {Xu}}, \bibinfo {author} {\bibfnamefont {N.}~\bibnamefont {Alidoust}},
  \bibinfo {author} {\bibfnamefont {I.}~\bibnamefont {Belopolski}}, \bibinfo
  {author} {\bibfnamefont {Z.}~\bibnamefont {Yuan}}, \bibinfo {author}
  {\bibfnamefont {G.}~\bibnamefont {Bian}}, \bibinfo {author} {\bibfnamefont
  {T.-R.}\ \bibnamefont {Chang}}, \bibinfo {author} {\bibfnamefont
  {H.}~\bibnamefont {Zheng}}, \bibinfo {author} {\bibfnamefont {V.~N.}\
  \bibnamefont {Strocov}}, \bibinfo {author} {\bibfnamefont {D.~S.}\
  \bibnamefont {Sanchez}}, \bibinfo {author} {\bibfnamefont {G.}~\bibnamefont
  {Chang}},  \emph {et~al.},\ }\href@noop {} {\bibfield  {journal} {\bibinfo
  {journal} {Nature Physics}\ }\textbf {\bibinfo {volume} {11}},\ \bibinfo
  {pages} {748} (\bibinfo {year} {2015}{\natexlab{b}})}\BibitemShut {NoStop}%
\bibitem [{\citenamefont {Yang}\ \emph {et~al.}(2015)\citenamefont {Yang},
  \citenamefont {Liu}, \citenamefont {Sun}, \citenamefont {Peng}, \citenamefont
  {Yang}, \citenamefont {Zhang}, \citenamefont {Zhou}, \citenamefont {Zhang},
  \citenamefont {Guo}, \citenamefont {Rahn} \emph {et~al.}}]{yang2015weyl}%
  \BibitemOpen
  \bibfield  {author} {\bibinfo {author} {\bibfnamefont {L.}~\bibnamefont
  {Yang}}, \bibinfo {author} {\bibfnamefont {Z.}~\bibnamefont {Liu}}, \bibinfo
  {author} {\bibfnamefont {Y.}~\bibnamefont {Sun}}, \bibinfo {author}
  {\bibfnamefont {H.}~\bibnamefont {Peng}}, \bibinfo {author} {\bibfnamefont
  {H.}~\bibnamefont {Yang}}, \bibinfo {author} {\bibfnamefont {T.}~\bibnamefont
  {Zhang}}, \bibinfo {author} {\bibfnamefont {B.}~\bibnamefont {Zhou}},
  \bibinfo {author} {\bibfnamefont {Y.}~\bibnamefont {Zhang}}, \bibinfo
  {author} {\bibfnamefont {Y.}~\bibnamefont {Guo}}, \bibinfo {author}
  {\bibfnamefont {M.}~\bibnamefont {Rahn}},  \emph {et~al.},\ }\href@noop {}
  {\bibfield  {journal} {\bibinfo  {journal} {Nature Physics}\ }\textbf
  {\bibinfo {volume} {11}},\ \bibinfo {pages} {728} (\bibinfo {year}
  {2015})}\BibitemShut {NoStop}%
\bibitem [{\citenamefont {Lv}\ \emph {et~al.}(2015{\natexlab{b}})\citenamefont
  {Lv}, \citenamefont {Xu}, \citenamefont {Weng}, \citenamefont {Ma},
  \citenamefont {Richard}, \citenamefont {Huang}, \citenamefont {Zhao},
  \citenamefont {Chen}, \citenamefont {Matt}, \citenamefont {Bisti} \emph
  {et~al.}}]{lv2015observation}%
  \BibitemOpen
  \bibfield  {author} {\bibinfo {author} {\bibfnamefont {B.}~\bibnamefont
  {Lv}}, \bibinfo {author} {\bibfnamefont {N.}~\bibnamefont {Xu}}, \bibinfo
  {author} {\bibfnamefont {H.}~\bibnamefont {Weng}}, \bibinfo {author}
  {\bibfnamefont {J.}~\bibnamefont {Ma}}, \bibinfo {author} {\bibfnamefont
  {P.}~\bibnamefont {Richard}}, \bibinfo {author} {\bibfnamefont
  {X.}~\bibnamefont {Huang}}, \bibinfo {author} {\bibfnamefont
  {L.}~\bibnamefont {Zhao}}, \bibinfo {author} {\bibfnamefont {G.}~\bibnamefont
  {Chen}}, \bibinfo {author} {\bibfnamefont {C.}~\bibnamefont {Matt}}, \bibinfo
  {author} {\bibfnamefont {F.}~\bibnamefont {Bisti}},  \emph {et~al.},\
  }\href@noop {} {\bibfield  {journal} {\bibinfo  {journal} {Nature Physics}\
  }\textbf {\bibinfo {volume} {11}},\ \bibinfo {pages} {724} (\bibinfo {year}
  {2015}{\natexlab{b}})}\BibitemShut {NoStop}%
\bibitem [{\citenamefont {Xu}\ \emph {et~al.}(2015{\natexlab{c}})\citenamefont
  {Xu}, \citenamefont {Weng}, \citenamefont {Lv}, \citenamefont {Matt},
  \citenamefont {Park}, \citenamefont {Bisti}, \citenamefont {Strocov},
  \citenamefont {Pomjakushina}, \citenamefont {Conder}, \citenamefont {Plumb}
  \emph {et~al.}}]{xu2015observation}%
  \BibitemOpen
  \bibfield  {author} {\bibinfo {author} {\bibfnamefont {N.}~\bibnamefont
  {Xu}}, \bibinfo {author} {\bibfnamefont {H.}~\bibnamefont {Weng}}, \bibinfo
  {author} {\bibfnamefont {B.}~\bibnamefont {Lv}}, \bibinfo {author}
  {\bibfnamefont {C.}~\bibnamefont {Matt}}, \bibinfo {author} {\bibfnamefont
  {J.}~\bibnamefont {Park}}, \bibinfo {author} {\bibfnamefont {F.}~\bibnamefont
  {Bisti}}, \bibinfo {author} {\bibfnamefont {V.}~\bibnamefont {Strocov}},
  \bibinfo {author} {\bibfnamefont {E.}~\bibnamefont {Pomjakushina}}, \bibinfo
  {author} {\bibfnamefont {K.}~\bibnamefont {Conder}}, \bibinfo {author}
  {\bibfnamefont {N.}~\bibnamefont {Plumb}},  \emph {et~al.},\ }\href@noop {}
  {\bibfield  {journal} {\bibinfo  {journal} {Preprint, arXiv:1507.03983}\ }
  (\bibinfo {year} {2015}{\natexlab{c}})}\BibitemShut {NoStop}%
\bibitem [{\citenamefont {Shekhar}\ \emph {et~al.}(2015)\citenamefont
  {Shekhar}, \citenamefont {Nayak}, \citenamefont {Sun}, \citenamefont
  {Schmidt}, \citenamefont {Nicklas}, \citenamefont {Leermakers}, \citenamefont
  {Zeitler}, \citenamefont {Schnelle}, \citenamefont {Grin}, \citenamefont
  {Felser} \emph {et~al.}}]{shekhar2015extremely}%
  \BibitemOpen
  \bibfield  {author} {\bibinfo {author} {\bibfnamefont {C.}~\bibnamefont
  {Shekhar}}, \bibinfo {author} {\bibfnamefont {A.~K.}\ \bibnamefont {Nayak}},
  \bibinfo {author} {\bibfnamefont {Y.}~\bibnamefont {Sun}}, \bibinfo {author}
  {\bibfnamefont {M.}~\bibnamefont {Schmidt}}, \bibinfo {author} {\bibfnamefont
  {M.}~\bibnamefont {Nicklas}}, \bibinfo {author} {\bibfnamefont
  {I.}~\bibnamefont {Leermakers}}, \bibinfo {author} {\bibfnamefont
  {U.}~\bibnamefont {Zeitler}}, \bibinfo {author} {\bibfnamefont
  {W.}~\bibnamefont {Schnelle}}, \bibinfo {author} {\bibfnamefont
  {J.}~\bibnamefont {Grin}}, \bibinfo {author} {\bibfnamefont {C.}~\bibnamefont
  {Felser}},  \emph {et~al.},\ }\href@noop {} {\bibfield  {journal} {\bibinfo
  {journal} {Nature Physics}\ }\textbf {\bibinfo {volume} {11}},\ \bibinfo
  {pages} {645} (\bibinfo {year} {2015})}\BibitemShut {NoStop}%
\bibitem [{\citenamefont {Zyuzin}\ and\ \citenamefont
  {Burkov}(2012)}]{zyuzin2012topological}%
  \BibitemOpen
  \bibfield  {author} {\bibinfo {author} {\bibfnamefont {A.}~\bibnamefont
  {Zyuzin}}\ and\ \bibinfo {author} {\bibfnamefont {A.}~\bibnamefont
  {Burkov}},\ }\href@noop {} {\bibfield  {journal} {\bibinfo  {journal} {Phys.
  Rev. B}\ }\textbf {\bibinfo {volume} {86}},\ \bibinfo {pages} {115133}
  (\bibinfo {year} {2012})}\BibitemShut {NoStop}%
\bibitem [{\citenamefont {Liu}\ \emph {et~al.}(2013)\citenamefont {Liu},
  \citenamefont {Ye},\ and\ \citenamefont {Qi}}]{liu2013chiral}%
  \BibitemOpen
  \bibfield  {author} {\bibinfo {author} {\bibfnamefont {C.-X.}\ \bibnamefont
  {Liu}}, \bibinfo {author} {\bibfnamefont {P.}~\bibnamefont {Ye}}, \ and\
  \bibinfo {author} {\bibfnamefont {X.-L.}\ \bibnamefont {Qi}},\ }\href@noop {}
  {\bibfield  {journal} {\bibinfo  {journal} {Phys. Rev. B}\ }\textbf {\bibinfo
  {volume} {87}},\ \bibinfo {pages} {235306} (\bibinfo {year}
  {2013})}\BibitemShut {NoStop}%
\bibitem [{\citenamefont {Zhang}\ \emph {et~al.}(2015)\citenamefont {Zhang},
  \citenamefont {Xu}, \citenamefont {Belopolski}, \citenamefont {Yuan},
  \citenamefont {Lin}, \citenamefont {Tong}, \citenamefont {Alidoust},
  \citenamefont {Lee}, \citenamefont {Huang}, \citenamefont {Lin} \emph
  {et~al.}}]{zhang2015observation}%
  \BibitemOpen
  \bibfield  {author} {\bibinfo {author} {\bibfnamefont {C.}~\bibnamefont
  {Zhang}}, \bibinfo {author} {\bibfnamefont {S.-Y.}\ \bibnamefont {Xu}},
  \bibinfo {author} {\bibfnamefont {I.}~\bibnamefont {Belopolski}}, \bibinfo
  {author} {\bibfnamefont {Z.}~\bibnamefont {Yuan}}, \bibinfo {author}
  {\bibfnamefont {Z.}~\bibnamefont {Lin}}, \bibinfo {author} {\bibfnamefont
  {B.}~\bibnamefont {Tong}}, \bibinfo {author} {\bibfnamefont {N.}~\bibnamefont
  {Alidoust}}, \bibinfo {author} {\bibfnamefont {C.-C.}\ \bibnamefont {Lee}},
  \bibinfo {author} {\bibfnamefont {S.-M.}\ \bibnamefont {Huang}}, \bibinfo
  {author} {\bibfnamefont {H.}~\bibnamefont {Lin}},  \emph {et~al.},\
  }\href@noop {} {\bibfield  {journal} {\bibinfo  {journal} {Nature Communications}}\ \textbf {\bibinfo {volume} {7}},\ \bibinfo {pages} {10735} (\bibinfo
  {year} {2016})}\BibitemShut {NoStop}%
\bibitem [{\citenamefont {Roy}\ and\ \citenamefont
  {Sau}(2015)}]{PhysRevB.92.125141}%
  \BibitemOpen
  \bibfield  {author} {\bibinfo {author} {\bibfnamefont {B.}~\bibnamefont
  {Roy}}\ and\ \bibinfo {author} {\bibfnamefont {J.~D.}\ \bibnamefont {Sau}},\
  }\href@noop {} {\bibfield  {journal} {\bibinfo  {journal} {Phys. Rev. B}\
  }\textbf {\bibinfo {volume} {92}},\ \bibinfo {pages} {125141} (\bibinfo
  {year} {2015})}\BibitemShut {NoStop}%
\bibitem [{\citenamefont {Bera}\ \emph {et~al.}(2015)\citenamefont {Bera},
  \citenamefont {Sau},\ and\ \citenamefont {Roy}}]{bera2015dirty}%
  \BibitemOpen
  \bibfield  {author} {\bibinfo {author} {\bibfnamefont {S.}~\bibnamefont
  {Bera}}, \bibinfo {author} {\bibfnamefont {J.~D.}\ \bibnamefont {Sau}}, \
  and\ \bibinfo {author} {\bibfnamefont {B.}~\bibnamefont {Roy}},\ }\href@noop
  {} {\bibfield  {journal} {\bibinfo  {journal} {Preprint, arXiv:1507.07551}\ }
  (\bibinfo {year} {2015})}\BibitemShut {NoStop}%
\bibitem [{\citenamefont {Trescher}\ \emph {et~al.}(2015)\citenamefont
  {Trescher}, \citenamefont {Sbierski}, \citenamefont {Brouwer},\ and\
  \citenamefont {Bergholtz}}]{Bergholtz2015}%
  \BibitemOpen
  \bibfield  {author} {\bibinfo {author} {\bibfnamefont {M.}~\bibnamefont
  {Trescher}}, \bibinfo {author} {\bibfnamefont {B.}~\bibnamefont {Sbierski}},
  \bibinfo {author} {\bibfnamefont {P.~W.}\ \bibnamefont {Brouwer}}, \ and\
  \bibinfo {author} {\bibfnamefont {E.~J.}\ \bibnamefont {Bergholtz}},\ }\href
  {\doibase 10.1103/PhysRevB.91.115135} {\bibfield  {journal} {\bibinfo
  {journal} {Phys. Rev. B}\ }\textbf {\bibinfo {volume} {91}},\ \bibinfo
  {pages} {115135} (\bibinfo {year} {2015})}\BibitemShut {NoStop}%
\bibitem [{\citenamefont {Soluyanov}\ \emph {et~al.}(2015)\citenamefont
  {Soluyanov}, \citenamefont {Gresch}, \citenamefont {Wang}, \citenamefont
  {Wu}, \citenamefont {Troyer}, \citenamefont {Dai},\ and\ \citenamefont
  {Bernevig}}]{soluyanov2015typeii}%
  \BibitemOpen
  \bibfield  {author} {\bibinfo {author} {\bibfnamefont {A.~A.}\ \bibnamefont
  {Soluyanov}}, \bibinfo {author} {\bibfnamefont {D.}~\bibnamefont {Gresch}},
  \bibinfo {author} {\bibfnamefont {Z.}~\bibnamefont {Wang}}, \bibinfo {author}
  {\bibfnamefont {Q.}~\bibnamefont {Wu}}, \bibinfo {author} {\bibfnamefont
  {M.}~\bibnamefont {Troyer}}, \bibinfo {author} {\bibfnamefont
  {M.}~\bibnamefont {Dai}}, \ and\ \bibinfo {author} {\bibfnamefont {B.~A.}\
  \bibnamefont {Bernevig}},\ }\href@noop {} {\bibfield  {journal} {\bibinfo
  {journal} {Nature}\ }\textbf {\bibinfo {volume} {527}},\ \bibinfo {pages}
  {495} (\bibinfo {year} {2015})}\BibitemShut {NoStop}%
\bibitem [{\citenamefont {Hosur}\ \emph {et~al.}(2012)\citenamefont {Hosur},
  \citenamefont {Parameswaran},\ and\ \citenamefont
  {Vishwanath}}]{hosur2012charge}%
  \BibitemOpen
  \bibfield  {author} {\bibinfo {author} {\bibfnamefont {P.}~\bibnamefont
  {Hosur}}, \bibinfo {author} {\bibfnamefont {S.}~\bibnamefont {Parameswaran}},
  \ and\ \bibinfo {author} {\bibfnamefont {A.}~\bibnamefont {Vishwanath}},\
  }\href@noop {} {\bibfield  {journal} {\bibinfo  {journal} {Phys. Rev. Lett.}\
  }\textbf {\bibinfo {volume} {108}},\ \bibinfo {pages} {046602} (\bibinfo
  {year} {2012})}\BibitemShut {NoStop}%
\bibitem [{\citenamefont {Altland}\ and\ \citenamefont
  {Simons}(2010)}]{Altland}%
  \BibitemOpen
  \bibfield  {author} {\bibinfo {author} {\bibfnamefont {A.}~\bibnamefont
  {Altland}}\ and\ \bibinfo {author} {\bibfnamefont {B.~D.}\ \bibnamefont
  {Simons}},\ }\href@noop {} {\emph {\bibinfo {title} {Condensed Matter Field
  Theory}}},\ Vol.~\bibinfo {volume} {2}\ (\bibinfo  {publisher} {Cambridge
  University Press},\ \bibinfo {year} {2010})\BibitemShut {NoStop}%
\bibitem [{\citenamefont {Gonzalez}\ \emph {et~al.}(1994)\citenamefont
  {Gonzalez}, \citenamefont {Guinea},\ and\ \citenamefont
  {Vozmediano}}]{gonzalez1994}%
  \BibitemOpen
  \bibfield  {author} {\bibinfo {author} {\bibfnamefont {J.}~\bibnamefont
  {Gonzalez}}, \bibinfo {author} {\bibfnamefont {F.}~\bibnamefont {Guinea}}, \
  and\ \bibinfo {author} {\bibfnamefont {V.~A.~M.}\ \bibnamefont
  {Vozmediano}},\ }\href@noop {} {\bibfield  {journal} {\bibinfo  {journal}
  {Nucl. Phys. B}\ }\textbf {\bibinfo {volume} {424}},\ \bibinfo {pages} {595}
  (\bibinfo {year} {1994})}\BibitemShut {NoStop}%
\bibitem [{\citenamefont {Son}(2007)}]{Son2007}%
  \BibitemOpen
  \bibfield  {author} {\bibinfo {author} {\bibfnamefont {D.~T.}\ \bibnamefont
  {Son}},\ }\href@noop {} {\bibfield  {journal} {\bibinfo  {journal} {Phys.
  Rev. B}\ }\textbf {\bibinfo {volume} {75}},\ \bibinfo {pages} {235423}
  (\bibinfo {year} {2007})}\BibitemShut {NoStop}%
\bibitem [{\citenamefont {Elias}\ \emph {et~al.}(2011)\citenamefont {Elias},
  \citenamefont {Gorbachev}, \citenamefont {Mayorov}, \citenamefont {Mo-rozov},
  \citenamefont {Zhukov}, \citenamefont {Blake}, \citenamefont {Ponomarenko},
  \citenamefont {Grigorieva}, \citenamefont {Novoselov}, \citenamefont
  {Guinea},\ and\ \citenamefont {Geim}}]{elias2011}%
  \BibitemOpen
  \bibfield  {author} {\bibinfo {author} {\bibfnamefont {D.~C.}\ \bibnamefont
  {Elias}}, \bibinfo {author} {\bibfnamefont {R.~V.}\ \bibnamefont
  {Gorbachev}}, \bibinfo {author} {\bibfnamefont {A.~S.}\ \bibnamefont
  {Mayorov}}, \bibinfo {author} {\bibfnamefont {S.~V.}\ \bibnamefont
  {Mo-rozov}}, \bibinfo {author} {\bibfnamefont {A.~A.}\ \bibnamefont
  {Zhukov}}, \bibinfo {author} {\bibfnamefont {P.}~\bibnamefont {Blake}},
  \bibinfo {author} {\bibfnamefont {L.~A.}\ \bibnamefont {Ponomarenko}},
  \bibinfo {author} {\bibfnamefont {I.~V.}\ \bibnamefont {Grigorieva}},
  \bibinfo {author} {\bibfnamefont {K.~S.}\ \bibnamefont {Novoselov}}, \bibinfo
  {author} {\bibfnamefont {F.}~\bibnamefont {Guinea}}, \ and\ \bibinfo {author}
  {\bibfnamefont {A.~K.}\ \bibnamefont {Geim}},\ }\href@noop {} {\bibfield
  {journal} {\bibinfo  {journal} {Nature Phys.}\ }\textbf {\bibinfo {volume}
  {7}},\ \bibinfo {pages} {701} (\bibinfo {year} {2011})}\BibitemShut {NoStop}%
\bibitem [{\citenamefont {Khveshchenko}(2001)}]{Khveshchenko2001}%
  \BibitemOpen
  \bibfield  {author} {\bibinfo {author} {\bibfnamefont {D.~V.}\ \bibnamefont
  {Khveshchenko}},\ }\href@noop {} {\bibfield  {journal} {\bibinfo  {journal}
  {Phys. Rev. Lett.}\ }\textbf {\bibinfo {volume} {87}},\ \bibinfo {pages}
  {246802} (\bibinfo {year} {2001})}\BibitemShut {NoStop}%
\bibitem [{\citenamefont {Drut}\ and\ \citenamefont
  {L\"ahde}(2009)}]{Drut2009}%
  \BibitemOpen
  \bibfield  {author} {\bibinfo {author} {\bibfnamefont {J.~E.}\ \bibnamefont
  {Drut}}\ and\ \bibinfo {author} {\bibfnamefont {T.~A.}\ \bibnamefont
  {L\"ahde}},\ }\href@noop {} {\bibfield  {journal} {\bibinfo  {journal} {Phys.
  Rev. Lett.}\ }\textbf {\bibinfo {volume} {102}},\ \bibinfo {pages} {026802}
  (\bibinfo {year} {2009})}\BibitemShut {NoStop}%
\bibitem [{\citenamefont {Burkov}\ and\ \citenamefont
  {Balents}(2011)}]{burkov2011weyl}%
  \BibitemOpen
  \bibfield  {author} {\bibinfo {author} {\bibfnamefont {A.}~\bibnamefont
  {Burkov}}\ and\ \bibinfo {author} {\bibfnamefont {L.}~\bibnamefont
  {Balents}},\ }\href@noop {} {\bibfield  {journal} {\bibinfo  {journal} {Phys.
  Rev. Lett.}\ }\textbf {\bibinfo {volume} {107}},\ \bibinfo {pages} {127205}
  (\bibinfo {year} {2011})}\BibitemShut {NoStop}%
\bibitem [{\citenamefont {Bulla}\ \emph {et~al.}(2008)\citenamefont {Bulla},
  \citenamefont {Costi},\ and\ \citenamefont {Pruschke}}]{ShankarRevModPhys}%
  \BibitemOpen
  \bibfield  {author} {\bibinfo {author} {\bibfnamefont {R.}~\bibnamefont
  {Bulla}}, \bibinfo {author} {\bibfnamefont {T.~A.}\ \bibnamefont {Costi}}, \
  and\ \bibinfo {author} {\bibfnamefont {T.}~\bibnamefont {Pruschke}},\
  }\href@noop {} {\bibfield  {journal} {\bibinfo  {journal} {Rev. Mod. Phys.}\
  }\textbf {\bibinfo {volume} {80}},\ \bibinfo {pages} {395} (\bibinfo {year}
  {2008})}\BibitemShut {NoStop}%
\bibitem [{\citenamefont {Barnes}\ \emph {et~al.}(2014)\citenamefont {Barnes},
  \citenamefont {Hwang}, \citenamefont {Throckmorton},\ and\ \citenamefont
  {Das~Sarma}}]{PhysRevB.89.235431}%
  \BibitemOpen
  \bibfield  {author} {\bibinfo {author} {\bibfnamefont {E.}~\bibnamefont
  {Barnes}}, \bibinfo {author} {\bibfnamefont {E.~H.}\ \bibnamefont {Hwang}},
  \bibinfo {author} {\bibfnamefont {R.~E.}\ \bibnamefont {Throckmorton}}, \
  and\ \bibinfo {author} {\bibfnamefont {S.}~\bibnamefont {Das~Sarma}},\ }\href
  {\doibase 10.1103/PhysRevB.89.235431} {\bibfield  {journal} {\bibinfo
  {journal} {Phys. Rev. B}\ }\textbf {\bibinfo {volume} {89}},\ \bibinfo
  {pages} {235431} (\bibinfo {year} {2014})}\BibitemShut {NoStop}%
\end{thebibliography}
\end{document}